\documentclass[conference]{IEEEtran}
\IEEEoverridecommandlockouts

\usepackage{url}
\usepackage{orcidlink}
\usepackage{graphics}

\usepackage[font={small},labelfont={bf}]{caption}

\usepackage{todonotes}

\usepackage{subcaption}

\usepackage{colortbl}

\usepackage{arydshln}
\definecolor{lightgray}{HTML}{F5F5F5}

\usepackage{enumitem}
\setlist[itemize]{leftmargin=*}
\setlist[enumerate]{leftmargin=*}
\usepackage{microtype}
\usepackage{balance}
\usepackage{multirow}

\usepackage{graphicx}

\makeatletter
\usepackage[compact]{titlesec}

\usepackage{listings}
\lstdefinestyle{bash}{language=bash,
    morekeywords={docker},
}
\usepackage[frozencache,cachedir=minted]{minted}
\usepackage{dblfloatfix}

\usepackage{cite}
\usepackage{amsmath,amssymb,amsfonts}
\usepackage{textcomp}
\usepackage{xcolor}
\usepackage[numbers]{natbib}
\def\BibTeX{{\rm B\kern-.05em{\sc i\kern-.025em b}\kern-.08em
    T\kern-.1667em\lower.7ex\hbox{E}\kern-.125emX}}
\usepackage{bibunits}
\defaultbibliographystyle{IEEEtranN}
\defaultbibliography{sample-base}

\usepackage{graphicx}
\usepackage{lipsum}
\usepackage{tcolorbox}
\usepackage[absolute,overlay]{textpos}

\newcommand\copyrighttext{%
    \footnotesize \textcopyright 2025 IEEE. Personal use of this material is permitted.  
    Permission from IEEE must be obtained for all other uses, in any current or future media, including reprinting/republishing this material for advertising or promotional purposes, creating new collective works, for resale or redistribution to servers or lists, or reuse of any copyrighted component of this work in other works.
    }
\newcommand\copyrightnotice{%
    \begin{tikzpicture}[remember picture,overlay]
        \node[anchor=south,yshift=10pt] at (current page.south) {\fbox{\parbox{\dimexpr\textwidth-\fboxsep-\fboxrule\relax}{\copyrighttext}}};
    \end{tikzpicture}%
}

\begin{document}
\title{WOW: Workflow-Aware Data Movement and Task Scheduling for Dynamic Scientific Workflows}

\author{
    \IEEEauthorblockN{
        Fabian Lehmann\orcidlink{0000-0003-0520-0792}\IEEEauthorrefmark{1}, 
        Jonathan Bader\orcidlink{0000-0003-0391-728X}\IEEEauthorrefmark{2}, 
        Friedrich Tschirpke\orcidlink{0000-0002-9376-2068}\IEEEauthorrefmark{1},\\
        Ninon De Mecquenem\orcidlink{0000-0003-3052-6129}\IEEEauthorrefmark{1}, 
        Ansgar Lößer\orcidlink{0000-0002-7627-9664}\IEEEauthorrefmark{3}, 
        Soeren Becker\orcidlink{0000-0001-6487-1268}\IEEEauthorrefmark{2},\\
        Katarzyna Ewa Lewińska\orcidlink{0000-0001-6560-1235}\IEEEauthorrefmark{1}\IEEEauthorrefmark{4}, 
        Lauritz Thamsen\orcidlink{0000-0003-3755-1503}\IEEEauthorrefmark{5}, 
        and Ulf Leser\orcidlink{0000-0003-2166-9582}\IEEEauthorrefmark{1}
    }
    \IEEEauthorblockA{\IEEEauthorrefmark{1}Humboldt-Universität zu Berlin, Germany,
        \IEEEauthorrefmark{2}Technische Universität Berlin, Germany,\\
        \IEEEauthorrefmark{3}Technische Universität Darmstadt, Germany,
        \IEEEauthorrefmark{4}University of Wisconsin--Madison, USA,
        \IEEEauthorrefmark{5}University of Glasgow, UK\\
        \{fabian.lehmann, tschirpf, mecquenn, leser\}@informatik.hu-berlin.de, jonathan.bader@tu-berlin.de,\\
        ansgar.loesser@kom.tu-darmstadt.de, soeren.becker@tu-berlin.de, lewinska@hu-berlin.de, lauritz.thamsen@glasgow.ac.uk
    }
}

\maketitle
\copyrightnotice

\begin{abstract}
Scientific workflows process extensive data sets over clusters of independent nodes, which requires a complex stack of infrastructure components, especially a resource manager (RM) for task-to-node assignment, a distributed file system (DFS) for data exchange between tasks, and a workflow engine to control task dependencies.
To enable a decoupled development and installation of these components, current architectures place intermediate data files during workflow execution independently of the future workload.
In data-intensive applications, this separation results in suboptimal schedules, as tasks are often assigned to nodes lacking input data, causing network traffic and bottlenecks.\looseness=-1

This paper presents WOW, a new scheduling approach for dynamic scientific workflow systems that steers both data movement and task scheduling to reduce network congestion and overall runtime.
For this, WOW creates speculative copies of intermediate files to prepare the execution of subsequently scheduled tasks.
WOW supports modern workflow systems that gain flexibility through the dynamic construction of execution plans.
We prototypically implemented WOW for the popular workflow engine Nextflow using Kubernetes as a resource manager.
In experiments with 16 synthetic and real workflows, WOW reduced makespan in all cases, with improvement of up to 94.5\% for workflow patterns and up to 53.2\% for real workflows, at a moderate increase of temporary storage space.
It also has favorable effects on CPU allocation and scales well with increasing cluster size.
\end{abstract}

\begin{bibunit}
\section{Introduction}
\begin{figure}
    \centering
    \includegraphics[width=0.85\columnwidth,trim={0.25mm 0.7mm 0.2mm 0.25mm},clip]{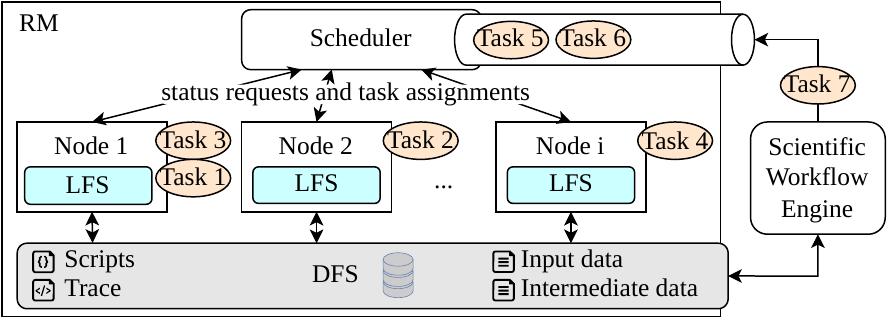}
    \caption{Overview of a cluster setup with a scientific workflow management system running on a resource manager. Nodes have a local file system (LFS) and access a distributed file system (DFS) to exchange data.}
    \label{fig:architectureOverviewOrig}
\end{figure}

Many scientific disciplines recently faced a steep increase in the size and amount of collected experimental data.
Such data is typically analyzed using a scientific workflow, which is a set of independently developed analysis programs (called tasks) connected through input/output relationships (called dependencies) managed by a scientific workflow engine.
For processing large data sets, these engines execute workflows on a distributed cluster by leveraging further infrastructure components~\cite{singh2019evaluating}.
The most important components are resource managers (RMs) to assign tasks whose input is available to nodes with free capacity and fitting resources, and distributed file systems (DFSs) to offer transparent access to data throughout the cluster (see Figure~\ref{fig:architectureOverviewOrig}).
Nextflow, Parsl, and Pegasus are examples of popular workflow engines.
Kubernetes and Slurm are widely used RMs.
HDFS, Ceph, and NFS are relevant DFSs.\looseness-1

Current architectures decouple the functionality of these three components to achieve a clear separation of concerns, with the known benefits for system development and maintenance~\cite{singh2019evaluating,costaCaseWorkflowAwareStorage2015a}.
However, this decoupling also yields significant disadvantages.
The disadvantage we focus on in this work is that the placement of intermediate data files created during workflow execution is determined independently of future workflow tasks.
The reason for this is that the different components pursue different optimization goals.
A DFS determines file locations with the goals of a balanced storage load across nodes and improved fault tolerance.
In contrast, a typical RM prioritizes the optimal distribution of the computational load of all ready tasks in the cluster.
Therein, the DFS is oblivious to future workloads, and the RM is oblivious to file locations.
Consequently, tasks read their input data from the DFS upon start-up and write their output to the DFS when finished.
This is not critical when clusters are equipped with high-speed, low-latency networks, such as InfiniBand, or if a workflow's runtime is heavily dominated by compute times. %
However, it leads to suboptimal schedules in data-intensive workflows executed over commodity networks, where data movement determines runtimes and, therefore, should be minimized as much as possible~\cite{ntakpeDataawareSimulationdrivenPlanning2022, sukhoroslovEfficientExecutionDataintensive2021,subediStackerAutonomicData2018,cimaHyperLoomPlatformDefining2018,delgadoTaskVineManagingInCluster2023,costaCaseWorkflowAwareStorage2015a}.

Consider a sequence of $n$ tasks where $t_1$ initiates the workflow by reading its input and tasks $t_2$ to $t_n$ each read output from their preceding tasks.
Typically, with a decentralized file system (DFS) independently managing data, all tasks fetch their inputs from the DFS and store their outputs in the DFS.
This often results in unnecessary data transfers.
For example, colocating $t_2$ with $t_1$ could retain $t_1$'s output on the same node, eliminating the need for data movement.
To achieve this, a central component must be able to steer both data placement and task scheduling in an intertwined manner, which is absent in current workflow systems.

Traditional workflow scheduling algorithms can jointly optimize data movement and task-node assignments but require that the entire workflow, task runtimes, and intermediate file sizes are known upfront~\cite{topcuogluPerformanceeffectiveLowcomplexityTask2002, sukhoroslovEfficientExecutionDataintensive2021,ntakpeDataawareSimulationdrivenPlanning2022,pietriSchedulingDataintensiveScientific2018}.
However, modern dynamic workflow engines create the workflow tasks dynamically based on the results of predecessor tasks.
Moreover, the systems treat tasks as black boxes with unpredictable resource requirements and output sizes.

In this paper, we propose WOW, a novel workflow-aware approach for dynamic, data-intensive scientific workflows on commodity clusters that efficiently intertwines data and task placement.
WOW features a three-step scheduler that connects decisions on data movement and task assignment such that (a) tasks are always assigned to nodes where all their input data is locally available (possibly after moving them there proactively using a speculative approach) and (b) data movement and task execution are performed in parallel (meaning that data possibly is moved before it is required).
To this end, the scheduler interacts with a dedicated data placement service (DPS), controlling all data movements during workflow execution to avoid network bottlenecks.
We implemented WOW for the popular combination of Nextflow workflows on Kubernetes clusters and evaluated it using 16 real-life workflows of varying complexity and from different scientific domains over two different DFSs, different commodity cluster sizes, and different network bandwidths.
Our experiments show that WOW achieves makespan improvements of up to 53.2\% for real-world workflows and 94.5\% for common workflow patterns.
WOW can thus significantly speed up the processing of data-intensive scientific workflows on commodity networks.\looseness-1

\thispagestyle{plain}
\pagestyle{plain}
\section{Background}
In this section, we examine how resource managers, distributed file systems, and workflow engines interact in current systems, leading to unnecessary network bottlenecks and longer execution times.\looseness-1

\subsection{Scientific Workflow Engine}
A scientific workflow engine comprises a workflow language and an execution engine~\cite{liu2020survey, singh2019evaluating, liuSurveyDataIntensiveScientific2015a}. 
Workflow languages define tasks and dependencies, where tasks are custom program binaries or scripts for data analysis, and dependencies represent input-output relationships.
The execution engine interprets the workflow iteratively, identifying ready tasks whose inputs are available and handing them to the RM's job queue for cluster execution.
The RM continuously signals back all tasks whose execution has finished successfully, which starts a new iteration within the workflow engine.
Execution stops when no more tasks are found to execute.
Together with the task, the engine must also inform the RM about the specific requirements of the task, especially the required main memory, CPU cores, and possibly access to special hardware like GPUs.
These requirements typically have to be provided by the user.
They are often inaccurate by large margins, making it impossible to use scheduling methods that rely on accurate knowledge of resource requirements~\cite{wittLearning2019}.
We denote the set of workflow tasks as $T = \{t_1, t_2, \ldots, t_i\}$.

There are two types of workflow engines: Static and dynamic.
Workflow engines for static workflows, such as Pegasus and Snakemake, know all physical tasks before the workflow execution starts.
In contrast, dynamic workflow engines such as Nextflow and Parsl support data-dependent language constructs~\cite{singh2019evaluating,liuSurveyDataIntensiveScientific2015a}.
This implies that the physical workflow tasks are only determined during execution and that they depend on the concrete workflow input.
Dynamic workflows make scheduling considerably more difficult because classical optimization approaches are not applicable.

\subsection{Resource Manager (RM)}\label{sec:resourceManager}
An RM manages a cluster of nodes and provides clients with access to its computing capacities.
In this work, the most important component of an RM is its job queue, a queue of tasks submitted by a client (the workflow engine) for execution in the cluster.
The RM continually monitors the cluster's node status.
When it detects a successfully finished task on a node, indicating available capacity, the RM schedules tasks from its queue to maximize cluster utilization while meeting individual resource requirements.
We denote the set of nodes in the cluster as $N = \{n_1,n_2,\ldots,n_j\}$, where $n_l$ is a tuple $(n^m_l,n^c_l)$, representing free memory ($n^m_l$) and the number of free cores ($n^c_l$).\looseness=-1

\subsection{Distributed File System (DFS)}
Workflow engines typically utilize a distributed file system (DFS) for data exchange between tasks, offering transparent file access across all cluster nodes (see Section~\ref{sub:relatedWork} for other approaches).
Clients read from and write to the DFS without knowing which node the accessed files actually reside in~\cite{costaCaseWorkflowAwareStorage2015a}.
The DFS determines a file's physical location and typically aims to achieve a fair share of storage and high fault tolerance by maintaining multiple (partial) file copies.
Workflow engines use a DFS by redirecting all file access from the tasks to it.
In this way, tasks read their input from the DFS and write their output to it.
This architecture is easy to use, highly portable across infrastructures, and provides DFS's fault tolerance capabilities to the workflow's intermediate results.
However, it implies that most task executions incur network traffic upon start-up and termination. %
Often, multiple tasks on different or even the same node read and write data to and from the DFS simultaneously.
This creates an accumulating load on the DFS and, thus, on the network.
In I/O-heavy workflows, where runtime is dominated by file reads/writes, this behavior quickly causes network bottlenecks and workflow slowdowns~\cite{delgadoTaskVineManagingInCluster2023,ntakpeDataawareSimulationdrivenPlanning2022,cimaHyperLoomPlatformDefining2018}.

\section{Our Approach}\label{sec:ourapproach}
This section describes WOW, a novel approach to workflow-aware data placement and scheduling for scientific workflows.

\subsection{Overview}
\begin{figure}
    \centering
    \includegraphics[width=0.85\columnwidth,trim={0.25mm 0.30mm 0.2mm 0.25mm},clip]{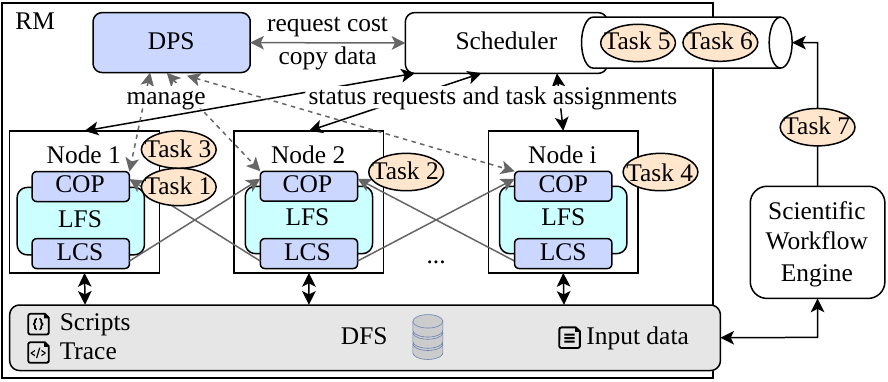}
    \caption{Overview of the cluster setup with our extensions highlighted in dark blue.
    The data placement service (DPS) keeps track of local files, calculates costs to start a task on a node, and manages data transfers.
    It uses local copy services (LCSs) to perform copy operations (COPs) that transfer intermediate data directly between nodes, bypassing the DFS.\looseness=-1}
    \label{fig:architectureOverviewOur}
\end{figure}

The general aim of WOW is to achieve a close integration of data placement and task scheduling by exploiting the information on the ready tasks provided by the workflow engine.
In a conventional setup, the RM places tasks on nodes based on their resource requirements.
The tasks then copy their required input from the DFS before they perform their computation or read data on demand during task execution and subsequently write their output to the DFS. %
This is a viable strategy for compute-heavy workflows like those that run on supercomputers but suboptimal for I/O-heavy scientific workflows analyzing large experimental data sets~\cite{liuSurveyDataIntensiveScientific2015a}.
To support data-intensive workflows, WOW places files and tasks so that tasks are always assigned to nodes where the required input data is already locally available (assuming that local storage is considerably faster than network access).
To this end, WOW primarily optimizes where to replicate files (once they were generated), followed by task assignment, rather than first placing tasks (once they are ready), followed by file movement.
File placement is controlled by a dedicated data placement service (DPS), which decides where and when to move task outputs.
Conceptually, this placement is performed through explicit copy operations (COPs) working in parallel to the execution of tasks, which allows for controlled and balanced network sharing.
WOW's high-level architecture is shown in Figure~\ref{fig:architectureOverviewOur}.\looseness-1

Note that WOW focuses on supporting the execution of workflows and thus only manages intermediate data, leaving the (precious) input data files in the DFS.
This intermediate data is only temporarily required as the output of one task and the input of at least one other task.
When all tasks needing an intermediate file have been completed, such a file could be deleted.
However, for fault tolerance, a minimal number of replicas could also be kept.

In the following, we describe two of WOW's three central components in more detail: the scheduler, which decides where to prepare and start task execution, and the DPS, which determines the placement of data files and their replicas in the cluster.
The third component is a local copy service (LCS) implemented by a daemon running on each node to perform the actual COPs when notified by the DPS.

\subsection{Scheduling Strategy}\label{sec::scheduler}
WOW implements a scheduling strategy that considers the co-location of tasks and data as the most important decision for reducing network I/O as much as possible.
A node $n_l$ with all data for a task $t_k$ is called \emph{prepared} for $t_k$.
The general idea of the scheduler is to try (a) moving data only when necessary and (b) placing data such that ready tasks can be started as soon as possible on a prepared node.
Following (a), WOW generally leaves all data produced by a task on the node where it was executed.
This data, however, is under the control of the DPS and copied to other nodes when the scheduler decides.
To implement (b), the scheduler assigns priorities to all ready tasks and follows an iterative process that continuously repeats three steps (details are given below):
\begin{enumerate}
    \item In the first step, ready tasks are assigned to a prepared node that is not at full capacity.
    This step will frequently place a task $t_{k+1}$ that is exclusively dependent on a task $t_k$ to the same node as $t_k$ because this node will have all the required input data.
    When this happens, no data is transferred over the network.
    Note that this step can only start tasks on a node prepared for this task; other tasks remain in the waiting queue.
    \item In the second step, unassigned ready tasks are strategically prepared on nodes that have enough available capacity to start the task as soon as all data is available.
    \item In contrast, the third step focuses on preparing high-priority tasks and creates COPs to speculatively prepare these tasks on nodes that do not yet have enough available resources.
\end{enumerate}
Once all three steps are processed, WOW waits until either a task finishes, a COP finishes, or a new task is submitted to the job queue.
Then, a new scheduling iteration starts.
Before describing the three steps in detail, we highlight a few features and possible pitfalls of our scheduling strategy.

First, our scheduler clearly is heuristic as global optimization is impossible in dynamic workflows, where the set of future tasks and the exact task runtimes and outputs are unknown at the time when scheduling decisions must be made.

Second, our strategy could, in principle, create a replica for each file on each node and start an arbitrary number of COPs from/to each node.
This would create network bottlenecks and excessive storage requirements.
We prevent both situations with two additional parameters.
First, we limit the number of parallel COPs for \emph{each node} by a threshold $c^{node}$.
A higher value for $c^{node}$ allows more parallel network operations, but each COP can only copy with $\frac{1}{c^{node}}$ of the maximum speed, leading to a later availability of all $c^{node}$ tasks which wastes CPU resources.
This was also shown in~\cite{donnellyConfugaScalableData2015}.
Second, we do not allow more than $c^{task}$ parallel COPs to prepare nodes for \emph{each task}.
Here, a higher value of $c^{task}$ results in more nodes that could start a given task, increasing the chance that a task is started earlier.
However, higher values of $c^{task}$ create more replicas and, thus, again, more network traffic.
Having a larger number of replicas also increases storage requirements.
However, these intermediate outputs only need to be stored temporarily. 
When all tasks requiring the data as input are finished, replicas can be deleted.\looseness-1

Third, our strategy logically connects data placement to task execution but effectively decouples these steps in time.
Effectively, even when the first step fully exhausts the node's resources reserved for task executions at a given time, the third step will, in parallel, already prepare nodes for the next tasks by initiating a constrained number of COPs.
This is different from previous approaches with independent placement decisions but task execution tightly coupled to the fetching of data.
That is, these approaches do not dissociate the steps in time, meaning that when a task is assigned to a node, it immediately starts downloading data to execute the task.

\textit{Task prioritization:}
The scheduler bases some of its decisions on a priority $t^p_k \in \mathbb{R}_{> 0}$ of each task $t_k$ that is calculated as soon as the task is submitted to the job queue.
Due to limited information on task runtime and successors, WOW uses a heuristic based on two metrics.
First, we consider a task's rank, which is the length of the path from the task to the sink in the abstract workflow graph.
The intuition is that we want to execute tasks with a higher rank early on because many other tasks depend on them.
As the second metric, we use the total size of the task's inputs; recall that tasks only become ready when all their inputs have been computed, meaning that these sizes are known.
The intuition here is that we want to schedule tasks earlier that require large inputs, as they usually need more time to finish (and thus increase the risk of being a straggler).
The final prioritization is done first by rank and second, in case of a tie, by input size.\looseness=-1

\textit{Step 1 ``Start ready tasks on prepared nodes'':}
The first step in WOW's scheduling strategy is to assign tasks to a prepared node for direct execution.
As there are often more ready tasks than resources, we select the tasks to run by solving a linear integer optimization problem over the set $T^{run} \subseteq T$ of all tasks where the task is prepared on at least one node and the set $N$ of all nodes with free resources.
A task $t_k$ is modeled as a quadruple $(t^m_k,t^c_k,N^{prep}_k,t^p_k)$, where $t^m_k$ is the amount of required memory and $t^c_k$ is the number of CPU cores requested.
$N^{prep}_k \subseteq {N}$ are all nodes with all input data for $t_k$ and $t^p_k$ is the task's priority as described above.
We create a binary matrix of task-node assignments $A^{k,l}$, where $a_{k,l} = 0$ if $n_l \notin N^{prep}_k$ or if the task $t_k$ is not going to be executed on $n_l$; otherwise $a_{k,l} = 1$.
The resulting optimization problem has the following constraints:
{%
\begin{align*}
    &\sum_{l=0}^{j} a_{k,l} &&\leq 1 && , k = 1,\ldots,i && \text{Execute task once} \\ 
    &\sum_{k=0}^{i} a_{k,l} * t^m_k &&\leq n^m_l && , l = 1,\ldots,j && \text{Memory constraint} \\ 
    &\sum_{k=0}^{i} a_{k,l} * t^c_k &&\leq n^c_l && , l = 1,\ldots,j && \text{CPU constraint}
\end{align*}}%
We maximize the system for: $\displaystyle \sum_{l=0}^{j}\sum_{k=0}^{i} a_{k,l} * t^p_k$;
the sum of the tasks' priorities assigned to nodes for execution.
The RM immediately starts these tasks on the respective nodes.

\textit{Step 2 ``Prepare ready tasks to fill available compute resources'':}
After step one, some nodes may still have free resources because there are no more ready tasks prepared for them.
In the second step, we, therefore, consider unassigned yet ready tasks and aim to prepare them on nodes with free compute capacities for the next scheduling iteration.
To this end, we sort all ready tasks in ascending order according to $|N^{prep}_k|$, i.e. we first select the tasks that are prepared on fewer nodes in the cluster.
Ties are resolved by the number of currently running COPs that prepare nodes for the task.
Going down this list of tasks, the scheduler communicates with the DPS to decide which task could start the earliest on a node with remaining compute resources; see Section~\ref{sec::filemanager} for details.
In its decision, the scheduler also ensures that $c^{node}$ and $c^{task}$ are not exceeded.

\textit{Step 3 ``Prepare high-priority tasks to use network capacity'':}
After step two, there can still be nodes with fewer COPs than $c^{node}$ because they work at full compute capacity and thus do not qualify for the COPs initiated in the second step.
In the third step, we leverage the free network capacity to start preparing further nodes for high-priority tasks.
First, we disregard tasks for which the maximum number $c^{task}$ of active COPs is already reached.
Then, we sort all the remaining tasks by priority.
Finally, the scheduler decides on which node to initiate COPs to prepare a task for future execution.
For this, it requests the cost of preparing a task on a node from the DPS and creates COPs for the task-node combination with the lowest cost.\looseness-1

\subsection{Data Placement Service (DPS)}\label{sec::filemanager}

The DPS keeps track of all files generated by a task and all file replicas.
Replicas are created only through explicit COPs.
COPs are the only operations that affect the network during workflow execution after we have consumed the workflow's input data from a DFS\footnote{We assume that tasks do not communicate to each other during execution, which is common in CPU-heavy simulation codes implemented with MPI, but very rare in the data-intensive applications we consider.
Furthermore, the DPS cannot control network operations resulting from activities in the cluster that are not related to the current workflow execution.
However, these are minimal in any sensible cluster setup.}.
The DPS thus centrally controls network traffic and the usage of network links to all nodes.
For each file, the DPS also stores its size and the task that created the file.

Several facts complicate the work of the DPS:
First, tasks may require multiple input files.
COPs that are initiated by the scheduler have a defined target node, but the DPS decides from which node to copy the replica to the target node.
Second, COPs (for large files) may take substantial time and may thus be active over multiple scheduling iterations.
The DPS's job is to deal with these complexities while ensuring that none of the network links to nodes in the cluster becomes overloaded and thus becomes a bottleneck.

The DPS achieves its goals by greedily solving an optimization problem at the request of the scheduler.
Recall that in steps two and three, the scheduler approaches the DPS with requests to evaluate the costs of task-node pairs, where the task is always the same, and the nodes stem from a set that is pre-selected by the scheduler according to the specific step.
For each request, the DPS returns an abstract price capturing the costs to prepare a task on a particular node.
The price should capture two components.
First, it should reflect the total amount of data transfer in the network, which will be lower for target nodes that already hold some of the task's inputs.
Second, the price should consider the maximal network load on each individual node, which will decrease when the required data movements are distributed over multiple source nodes.
Both measures should be minimized but contradict each other; we give equal weight to both aspects.

We use a greedy heuristic to determine the best configuration and its price.
The DPS first sorts all files missing on the target node by their size.
For each of these, it identifies all the source nodes that hold a replica of this file and chooses the one that has the lowest load already assigned for this COP.
For the first element of the list of files, this only leads to ties, as no load has been assigned yet; these are resolved randomly.
But further down the list, the number of ties will decrease and the assignments will become more directed.
Once all assignments are made, their total price is computed as the weighted sum of the total network traffic incurred and the maximal load of a participating node.

Clearly, this strategy implies that decisions on COPs are taken individually for each task, which does not guarantee the achievement of an optimal solution.
However, to get closer to an optimum, it would be necessary to collect COPs over some period of time and then decide on them all at once.
This, in turn, would mean that the scheduler needs to postpone its decisions for some time, leading to idle networks and idle nodes.\looseness-1

\section{Prototype}\label{sec:prototype}
To demonstrate the benefits of our workflow-aware data placement approach for real workflow executions, we implemented a prototype for the combination of Nextflow (as workflow engine) and Kubernetes (as resource manager). 
Here, we describe how we built on an existing workflow-scheduler API, adapted Nextflow, realized our scheduler and data placement service, and implemented the data transfers.
We publish all source code on GitHub for reproducibility\footnote{\label{foot::github}
\url{https://github.com/WOW-WorkflowScheduler/}}.
\subsection{Common Workflow Scheduler}

To have workflow information available for placing tasks within Kubernetes, we implemented WOW based on the Common Workflow Scheduler (CWS)~\cite{lehmannHowWorkflowEngines2023}.
CWS is a general interface between workflow engines and resource managers, with a prototype integrating Nextflow and Kubernetes.
The CWS allows the abstract DAG to be passed to WOW, which enables us to prioritize specific tasks rather than performing a first-in-first-out (FIFO) scheduling.
In addition, the CWS provides information about the files a task requires, which is essential for our approach.

For our prototype, we integrate the DPS capabilities directly into the Common Workflow Scheduler, ensuring seamless interaction between the scheduler and the DPS.

\subsection{Implementation with Nextflow}

In a Nextflow-Kubernetes setup, Nextflow operates in a Kubernetes pod, using a DFS for data exchange between tasks.
Each task runs in its own pod, accessing the same file system.

Nextflow creates a working folder within the DFS, with a subfolder for each task that contains a specific task script and a wrapper script that creates a temporary folder on the executing machine's local file system (LFS). 
The input data is linked to this folder, and the task script is executed within it.
After task execution, the wrapper script copies all outputs to the task folder on the DFS, where task statistics and traces are also written.
Once a task pod is completed, Nextflow scans all new files in the task folder and determines which tasks now have all their required inputs and are ready to run.

We made two major changes to Nextflow to implement WOW.
To allow tasks to read and write data locally on the node, we mount a local path and modify the wrapper script to read and write input and output files from the local path instead of the DFS.
Also, to track a task's output, we extend Nextflow's wrapper script to write a list of output files to the DFS.
In this list, we store the full path for each file, and if it is a symlink, we also store the target path.
After the task is finished, Nextflow scans the list of output files instead of the DFS to determine the ready tasks.
We catch any file access to the output and, therefore, patch Nextflow methods and classes to implement the following:
\begin{itemize}
    \item When metadata is requested for a file, we provide the information from our file list in the DFS.
    \item When the file's content is accessed, we request the location of one file's replica from the scheduler and then read it from the Local Copy Service (see Section \ref{sec::lcs}).
    \item When a file is manipulated, we copy it to the Nextflow pod's node and then modify it locally.
    After a file is manipulated, we notify the scheduler.
    The scheduler then sets the only valid file location to the node where the Nextflow pod is running and invalidates all other locations.
\end{itemize}

\subsection{Scheduler with Data Placement Service}

In the scheduler, we have implemented our three-step scheduling strategy described in Section~\ref{sec:ourapproach}.
The linear integer problem of the first step is solved using Google's OR Tools\footnote{\url{https://developers.google.com/optimization}}.
We terminate the optimizer after ten seconds if no optimum is found and use the best solution found so far.
In our experiments, this threshold did not apply at all, as WOW always found the optimum in less than two seconds, with a median optimization time of 11 ms and the 99th percentile at 112 ms.
For the second step, we approximate the transfer time before a task can start by the sum of the bytes to copy.

The scheduler communicates with the LCS to start COPs.
Therefore, we keep track of which LCS is on which node.
We extend the CWS to expose this information to Nextflow.
The scheduler also tracks file locations, replicas, ongoing and planned copies.
COPs are atomic units that always consider the full set of file replicas needed for task preparation, not individual files.
When a COP finishes, all its created file replicas are added to the record; none are added upon COP failure.\looseness=-1

\subsection{Local Copy Service (LCS)}\label{sec::lcs}

We use FTP as the data transmission protocol for the LCS and start a small LCS daemon that exposes the local storage via FTP and runs the COPs on all nodes.
We reuse the LCS to run COPs, as starting a new service for every COP would introduce considerable overhead.
This is especially true for workflow tasks that run only for a few seconds, as the LCS startup time could otherwise double their execution time.

We keep the workflow input data in the DFS, as this allows us to change only how the workflow is executed, not how the input data is provided, thus increasing portability.
That is, we only use the LCS to manage intermediate data, not initial input data.\looseness-1

\section{Evaluation Setup}

This section describes the 16 workflows, the execution environment, and the experimental design that we used to evaluate WOW within our prototype.

\subsection{Workflows and Workflow Patterns}
We evaluate WOW using 16 different workflows, which fall into three classes: real-world workflows (4), synthetic workflows (7), and synthetic workflow patterns (5). 
The details can be found in Table~\ref{tab:workflow_input}.
We consider the real-life workflows to be our main result, while the synthetic workflows are used to complement the breadth of evaluation, and the patterns serve to showcase specific workflow topologies and how WOW handles them.\looseness-1

\begin{table}
\centering
\caption{Evaluation workflows, including real-world workflows, synthetic workflows, and workflow patterns. The ``Inputs in GB'' is the sum of the data set used for the particular workflow. The ``Generated GB'' is the sum of all data generated by the workflow, either as temporary, intermediate or output data. The ``Factor'' is the factor between the input data and the generated data. Abstract tasks are the logical steps of a workflow, while physical tasks are the concrete instances.}
\footnotesize
\begin{tabular}{p{0.05cm}|l|r|r|r|r|r}
& \multicolumn{1}{p{1.2cm}|}{\parbox[c][2.5em]{\hsize}{\centering Workflow}} & \multicolumn{1}{p{0.67cm}|}{\parbox[c][3.3em]{\hsize}{\centering Inputs in GB}} & \multicolumn{1}{p{1.05cm}|}{\parbox[c][2.5em]{\hsize}{\centering Generated GB}} & \multicolumn{1}{p{0.4cm}|}{\parbox[c][2.5em]{\hsize}{\centering Fac-tor}} & \multicolumn{1}{p{0.85cm}|}{\parbox[c][2.5em]{\hsize}{\centering Abstract tasks}} & \multicolumn{1}{p{0.8cm}}{\parbox[c][2.5em]{\hsize}{\centering Physical tasks}} \\
\hline
\multirow{4}{*}{\cellcolor{white}\rotatebox{90}{Real-World}} & RNA-Seq & 139.1 & 598.3 & 4.3 & 53 & 1,269 \\
 & \cellcolor{lightgray}Sarek & \cellcolor{lightgray}205.9 & \cellcolor{lightgray}918.8 & \cellcolor{lightgray}4.5 & \cellcolor{lightgray}49 & \cellcolor{lightgray}8,656 \\
 & Chip-Seq & 141.2 & 787.2 & 5.6 & 48 & 3,537 \\
 & \cellcolor{lightgray}Rangeland & \cellcolor{lightgray}303.2 & \cellcolor{lightgray}274.0 & \cellcolor{lightgray}0.9 & \cellcolor{lightgray}8 & \cellcolor{lightgray}3,184 \\
\hdashline
\multirow{7}{*}{\cellcolor{white}\rotatebox{90}{Synthetic}} & Syn. BLAST & 21.9 & 151.0 & 6.9 & 4 & 198 \\
 & \cellcolor{lightgray}Syn. BWA & \cellcolor{lightgray}19.4 & \cellcolor{lightgray}152.8 & \cellcolor{lightgray}7.9 & \cellcolor{lightgray}5 & \cellcolor{lightgray}198 \\
 & Syn. Cycles & 20.4 & 157.9 & 7.7 & 7 & 198 \\
 & \cellcolor{lightgray}Syn. Genome & \cellcolor{lightgray}21.9 & \cellcolor{lightgray}154.7 & \cellcolor{lightgray}7.1 & \cellcolor{lightgray}5 & \cellcolor{lightgray}198 \\
 & Syn. Montage & 19.8 & 168.8 & 8.5 & 8 & 198 \\
 & \cellcolor{lightgray}Syn. Seismology & \cellcolor{lightgray}20.7 & \cellcolor{lightgray}150.7 & \cellcolor{lightgray}7.3 & \cellcolor{lightgray}2 & \cellcolor{lightgray}198 \\
 & Syn. Soykb & 22.3 & 160.0 & 7.2 & 14 & 196 \\
\hdashline
\cellcolor{lightgray}\multirow{5}{*}{\cellcolor{white}\rotatebox{90}{Pattern}} & \cellcolor{lightgray}All in One & \cellcolor{lightgray}0.0 & \cellcolor{lightgray}180.3 & \cellcolor{lightgray}- & \cellcolor{lightgray}2 & \cellcolor{lightgray}101 \\
 & Chain & 0.0 & 180.3 & - & 2 & 200 \\
 & \cellcolor{lightgray}Fork & \cellcolor{lightgray}0.0 & \cellcolor{lightgray}99.4 & \cellcolor{lightgray}- & \cellcolor{lightgray}2 & \cellcolor{lightgray}101 \\
 & Group & 0.0 & 180.3 & - & 2 & 134 \\
 & \cellcolor{lightgray}Group Multiple & \cellcolor{lightgray}0.0 & \cellcolor{lightgray}270.5 & \cellcolor{lightgray}- & \cellcolor{lightgray}3 & \cellcolor{lightgray}160 \\
\end{tabular}
\label{tab:workflow_input}
\end{table}

\begin{figure}
    \centering
    \begin{subfigure}{.27\columnwidth}
      \centering
      \includegraphics[width=.675\columnwidth,clip]{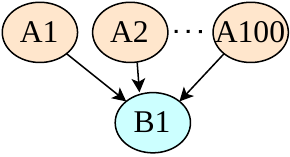}
      \caption{All in One}
	  \label{fig:allIntoOne}
    \end{subfigure}%
    \begin{subfigure}{.27\columnwidth}
      \centering
      \includegraphics[width=.675\columnwidth,clip]{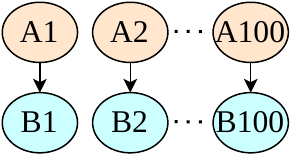}
      \caption{Chain}
	  \label{fig:chain}
    \end{subfigure}%
    \begin{subfigure}{.27\columnwidth}
      \centering
      \includegraphics[width=.675\columnwidth,clip]{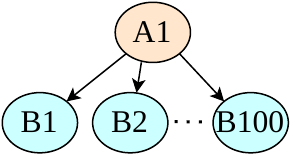}
      \caption{Fork}
	  \label{fig:fork}
    \end{subfigure}%
    \hfill
    \begin{subfigure}{.43\columnwidth}
      \centering
      \includegraphics[width=.9\columnwidth,clip]{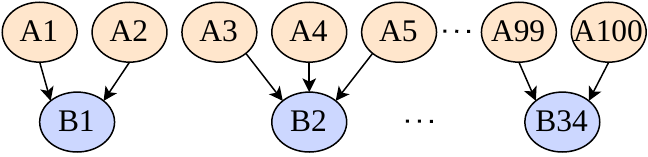}
      \caption{Group}
	  \label{fig:group}
    \end{subfigure}%
    \begin{subfigure}{.43\columnwidth}
      \centering
      \vspace{.2cm}
      \includegraphics[width=.9\columnwidth,clip]{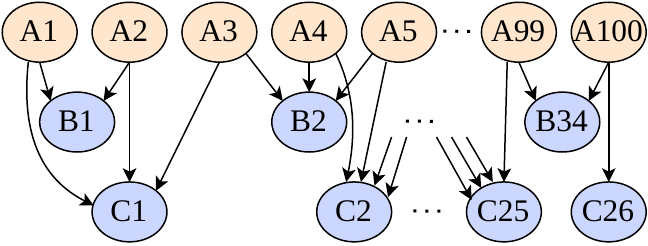}
      \caption{Group Multiple}
	  \label{fig:groupMultiple}
    \end{subfigure}%
    \caption{Five evaluated workflow patterns consisting of the Tasks A, B, and C. Task A always writes a random file ranging from 0.8 to 1~GB. Task B and Task C read all inputs and merge them into a single file.}
    \label{fig:workflowPatterns}
\end{figure}%
\begingroup
\setlength{\tabcolsep}{1.25mm}
\begin{table*}
\centering
\caption{Results for workflow execution using Ceph and NFS with eight nodes and 1 Gbit network speed. The original median makespan of Nextflow's original scheduling (``Orig'') is given in minutes, and the median change in makespan for the CWS and our WOW approach is compared to the original in percent. The median allocated CPU hours for the original execution are measured as the sum of tasks' runtimes multiplied by the number of allocated CPUs over all tasks. As with makespan, we report the relative change for CWS and WOW over the original. Finally, we show the median of how many of the tasks ran without needing any COPs because all required data was local (``none'') and the median of how many of the COPs transferred data that was used by tasks (``used''). A greener background indicates better values and a redder background indicates worse values.\looseness=-1}
\footnotesize
\resizebox{\textwidth}{!}{
\begin{tabular}{c|l|r|r|r||r|r|r||r|r||r|r|r||r|r|r||r|r|r|}
 & \multicolumn{1}{c|}{} & \multicolumn{8}{c||}{Ceph} & \multicolumn{8}{c|}{NFS}\\
 & \multicolumn{1}{c|}{Workflow} & \multicolumn{3}{c||}{Makespan [min]} & \multicolumn{3}{c||}{CPU allocated [h]} & \multicolumn{2}{c||}{WOW COPs} & \multicolumn{3}{c||}{Makespan [min]} & \multicolumn{3}{c||}{CPU allocated [h]} & \multicolumn{2}{c|}{WOW COPs}\\
 &  & Orig & CWS & WOW & Orig & CWS & WOW & none & used & Orig & CWS & WOW & Orig & CWS & WOW & none & used\\
\hline
\multirow{4}{*}{\cellcolor{white}\rotatebox{90}{Real-World}} & RNA-Seq & 181.1 & \cellcolor[RGB]{236,255,236} -6.1\% & \cellcolor[RGB]{230,255,230} -18.3\% & 474.4 & \cellcolor[RGB]{255,236,236} 7.9\% & \cellcolor[RGB]{233,255,233} -13.6\% & \cellcolor[RGB]{194,255,194} 90.1\% & \cellcolor[RGB]{225,255,225} 28.8\% & 413.0 & \cellcolor[RGB]{238,255,238} -2.6\% & \cellcolor[RGB]{213,255,213} -53.2\% & 1,354.6 & \cellcolor[RGB]{255,233,233} 12.4\% & \cellcolor[RGB]{209,255,209} -60.8\% & \cellcolor[RGB]{195,255,195} 89.1\% & \cellcolor[RGB]{228,255,228} 22.9\% \\
 & \cellcolor{lightgray}Sarek & \cellcolor{lightgray}305.0 & \cellcolor{lightgray}\cellcolor[RGB]{236,255,236} -7.0\% & \cellcolor{lightgray}\cellcolor[RGB]{237,255,237} -4.2\% & \cellcolor{lightgray}997.9 & \cellcolor{lightgray}\cellcolor[RGB]{255,239,239} 1.2\% & \cellcolor{lightgray}\cellcolor[RGB]{237,255,237} -4.1\% & \cellcolor{lightgray}\cellcolor[RGB]{197,255,197} 86.0\% & \cellcolor{lightgray}\cellcolor[RGB]{219,255,219} 40.4\% & \cellcolor{lightgray}557.5 & \cellcolor{lightgray}\cellcolor[RGB]{239,255,239} -1.3\% & \cellcolor{lightgray}\cellcolor[RGB]{218,255,218} -42.6\% & \cellcolor{lightgray}1,806.1 & \cellcolor{lightgray}\cellcolor[RGB]{239,255,239} -0.4\% & \cellcolor{lightgray}\cellcolor[RGB]{219,255,219} -40.4\% & \cellcolor{lightgray}\cellcolor[RGB]{197,255,197} 85.8\% & \cellcolor{lightgray}\cellcolor[RGB]{220,255,220} 39.1\% \\
 & Chip-Seq & 221.1 & \cellcolor[RGB]{255,237,237} 4.9\% & \cellcolor[RGB]{232,255,232} -15.4\% & 710.3 & \cellcolor[RGB]{255,236,236} 7.1\% & \cellcolor[RGB]{237,255,237} -6.0\% & \cellcolor[RGB]{200,255,200} 79.6\% & \cellcolor[RGB]{230,255,230} 18.6\% & 375.0 & \cellcolor[RGB]{255,235,235} 9.6\% & \cellcolor[RGB]{217,255,217} -44.8\% & 1,293.8 & \cellcolor[RGB]{255,232,232} 15.7\% & \cellcolor[RGB]{218,255,218} -42.3\% & \cellcolor[RGB]{200,255,200} 78.9\% & \cellcolor[RGB]{231,255,231} 17.7\% \\
 & \cellcolor{lightgray}Rangeland & \cellcolor{lightgray}166.0 & \cellcolor{lightgray}\cellcolor[RGB]{239,255,239} -1.9\% & \cellcolor{lightgray}\cellcolor[RGB]{237,255,237} -4.3\% & \cellcolor{lightgray}468.3 & \cellcolor{lightgray}\cellcolor[RGB]{255,239,239} 0.1\% & \cellcolor{lightgray}\cellcolor[RGB]{238,255,238} -2.1\% & \cellcolor{lightgray}\cellcolor[RGB]{195,255,195} 88.7\% & \cellcolor{lightgray}\cellcolor[RGB]{209,255,209} 60.7\% & \cellcolor{lightgray}181.2 & \cellcolor{lightgray}\cellcolor[RGB]{239,255,239} -0.7\% & \cellcolor{lightgray}\cellcolor[RGB]{233,255,233} -13.4\% & \cellcolor{lightgray}476.2 & \cellcolor{lightgray}\cellcolor[RGB]{255,239,239} 0.7\% & \cellcolor{lightgray}\cellcolor[RGB]{238,255,238} -2.8\% & \cellcolor{lightgray}\cellcolor[RGB]{195,255,195} 88.7\% & \cellcolor{lightgray}\cellcolor[RGB]{207,255,207} 64.3\% \\
\hdashline
\multirow{7}{*}{\cellcolor{white}\rotatebox{90}{Synthetic}} & Syn. BLAST & 35.0 & \cellcolor[RGB]{255,239,239} 0.5\% & \cellcolor[RGB]{219,255,219} -41.6\% & 1.5 & \cellcolor[RGB]{233,255,233} -12.9\% & \cellcolor[RGB]{214,255,214} -51.3\% & \cellcolor[RGB]{190,255,190} 99.0\% & \cellcolor[RGB]{230,255,230} 18.2\% & 55.6 & \cellcolor[RGB]{255,239,239} 0.7\% & \cellcolor[RGB]{209,255,209} -60.8\% & 3.9 & \cellcolor[RGB]{236,255,236} -6.2\% & \cellcolor[RGB]{208,255,208} -62.6\% & \cellcolor[RGB]{190,255,190} 99.0\% & \cellcolor[RGB]{230,255,230} 18.2\% \\
 & \cellcolor{lightgray}Syn. BWA & \cellcolor{lightgray}37.7 & \cellcolor{lightgray}\cellcolor[RGB]{239,255,239} -1.0\% & \cellcolor{lightgray}\cellcolor[RGB]{209,255,209} -61.1\% & \cellcolor{lightgray}12.0 & \cellcolor{lightgray}\cellcolor[RGB]{239,255,239} -0.6\% & \cellcolor{lightgray}\cellcolor[RGB]{200,255,200} -78.8\% & \cellcolor{lightgray}\cellcolor[RGB]{190,255,190} 100.0\% & \cellcolor{lightgray}\cellcolor[RGB]{255,255,255} 0.0\% & \cellcolor{lightgray}81.7 & \cellcolor{lightgray}\cellcolor[RGB]{255,239,239} 1.2\% & \cellcolor{lightgray}\cellcolor[RGB]{198,255,198} -82.7\% & \cellcolor{lightgray}65.7 & \cellcolor{lightgray}\cellcolor[RGB]{239,255,239} -0.9\% & \cellcolor{lightgray}\cellcolor[RGB]{191,255,191} -96.0\% & \cellcolor{lightgray}\cellcolor[RGB]{190,255,190} 100.0\% & \cellcolor{lightgray}\cellcolor[RGB]{255,255,255} 0.0\% \\
 & Syn. Cycles & 20.0 & \cellcolor[RGB]{255,238,238} 3.6\% & \cellcolor[RGB]{216,255,216} -47.9\% & 9.2 & \cellcolor[RGB]{237,255,237} -4.2\% & \cellcolor[RGB]{200,255,200} -78.4\% & \cellcolor[RGB]{195,255,195} 89.9\% & \cellcolor[RGB]{214,255,214} 51.1\% & 55.6 & \cellcolor[RGB]{238,255,238} -2.8\% & \cellcolor[RGB]{199,255,199} -81.3\% & 30.9 & \cellcolor[RGB]{236,255,236} -6.9\% & \cellcolor[RGB]{194,255,194} -91.3\% & \cellcolor[RGB]{195,255,195} 89.4\% & \cellcolor[RGB]{215,255,215} 48.8\% \\
 & \cellcolor{lightgray}Syn. Genome & \cellcolor{lightgray}28.6 & \cellcolor{lightgray}\cellcolor[RGB]{237,255,237} -4.7\% & \cellcolor{lightgray}\cellcolor[RGB]{209,255,209} -62.0\% & \cellcolor{lightgray}22.8 & \cellcolor{lightgray}\cellcolor[RGB]{255,239,239} 0.5\% & \cellcolor{lightgray}\cellcolor[RGB]{195,255,195} -88.7\% & \cellcolor{lightgray}\cellcolor[RGB]{193,255,193} 92.9\% & \cellcolor{lightgray}\cellcolor[RGB]{213,255,213} 53.8\% & \cellcolor{lightgray}92.9 & \cellcolor{lightgray}\cellcolor[RGB]{255,239,239} 0.7\% & \cellcolor{lightgray}\cellcolor[RGB]{196,255,196} -86.3\% & \cellcolor{lightgray}100.2 & \cellcolor{lightgray}\cellcolor[RGB]{255,238,238} 3.3\% & \cellcolor{lightgray}\cellcolor[RGB]{192,255,192} -95.0\% & \cellcolor{lightgray}\cellcolor[RGB]{193,255,193} 92.9\% & \cellcolor{lightgray}\cellcolor[RGB]{216,255,216} 47.8\% \\
 & Syn. Montage & 31.4 & \cellcolor[RGB]{238,255,238} -2.8\% & \cellcolor[RGB]{217,255,217} -44.6\% & 7.4 & \cellcolor[RGB]{255,233,233} 13.2\% & \cellcolor[RGB]{195,255,195} -88.8\% & \cellcolor[RGB]{209,255,209} 61.6\% & \cellcolor[RGB]{202,255,202} 74.0\% & 85.8 & \cellcolor[RGB]{238,255,238} -2.0\% & \cellcolor[RGB]{200,255,200} -78.7\% & 27.2 & \cellcolor[RGB]{255,236,236} 6.8\% & \cellcolor[RGB]{193,255,193} -93.0\% & \cellcolor[RGB]{209,255,209} 61.1\% & \cellcolor[RGB]{203,255,203} 72.6\% \\
 & \cellcolor{lightgray}Syn. Seismology & \cellcolor{lightgray}31.4 & \cellcolor{lightgray}\cellcolor[RGB]{255,239,239} 0.8\% & \cellcolor{lightgray}\cellcolor[RGB]{229,255,229} -20.9\% & \cellcolor{lightgray}1.5 & \cellcolor{lightgray}\cellcolor[RGB]{232,255,232} -15.9\% & \cellcolor{lightgray}\cellcolor[RGB]{218,255,218} -42.6\% & \cellcolor{lightgray}\cellcolor[RGB]{190,255,190} 99.5\% & \cellcolor{lightgray}\cellcolor[RGB]{215,255,215} 50.0\% & \cellcolor{lightgray}45.5 & \cellcolor{lightgray}\cellcolor[RGB]{255,239,239} 0.5\% & \cellcolor{lightgray}\cellcolor[RGB]{216,255,216} -47.4\% & \cellcolor{lightgray}2.4 & \cellcolor{lightgray}\cellcolor[RGB]{239,255,239} -1.6\% & \cellcolor{lightgray}\cellcolor[RGB]{222,255,222} -35.5\% & \cellcolor{lightgray}\cellcolor[RGB]{190,255,190} 99.5\% & \cellcolor{lightgray}\cellcolor[RGB]{215,255,215} 50.0\% \\
 & Syn. Soykb & 31.6 & \cellcolor[RGB]{237,255,237} -4.0\% & \cellcolor[RGB]{211,255,211} -56.9\% & 7.9 & \cellcolor[RGB]{255,233,233} 13.2\% & \cellcolor[RGB]{202,255,202} -74.8\% & \cellcolor[RGB]{194,255,194} 91.8\% & \cellcolor[RGB]{210,255,210} 58.5\% & 65.7 & \cellcolor[RGB]{239,255,239} -1.4\% & \cellcolor[RGB]{203,255,203} -72.9\% & 29.3 & \cellcolor[RGB]{238,255,238} -3.3\% & \cellcolor[RGB]{193,255,193} -92.6\% & \cellcolor[RGB]{196,255,196} 87.8\% & \cellcolor[RGB]{213,255,213} 53.3\% \\
\hdashline
\cellcolor{lightgray}\multirow{5}{*}{\cellcolor{white}\rotatebox{90}{Pattern}} & \cellcolor{lightgray}All in One & \cellcolor{lightgray}32.5 & \cellcolor{lightgray}\cellcolor[RGB]{238,255,238} -2.8\% & \cellcolor{lightgray}\cellcolor[RGB]{215,255,215} -49.3\% & \cellcolor{lightgray}2.5 & \cellcolor{lightgray}\cellcolor[RGB]{237,255,237} -5.2\% & \cellcolor{lightgray}\cellcolor[RGB]{205,255,205} -69.1\% & \cellcolor{lightgray}\cellcolor[RGB]{190,255,190} 99.0\% & \cellcolor{lightgray}\cellcolor[RGB]{215,255,215} 50.0\% & \cellcolor{lightgray}40.6 & \cellcolor{lightgray}\cellcolor[RGB]{255,239,239} 0.1\% & \cellcolor{lightgray}\cellcolor[RGB]{209,255,209} -60.1\% & \cellcolor{lightgray}2.7 & \cellcolor{lightgray}\cellcolor[RGB]{255,238,238} 2.5\% & \cellcolor{lightgray}\cellcolor[RGB]{204,255,204} -71.4\% & \cellcolor{lightgray}\cellcolor[RGB]{190,255,190} 99.0\% & \cellcolor{lightgray}\cellcolor[RGB]{215,255,215} 50.0\% \\
 & Chain & 16.2 & \cellcolor[RGB]{255,238,238} 2.8\% & \cellcolor[RGB]{196,255,196} -86.4\% & 40.3 & \cellcolor[RGB]{233,255,233} -13.4\% & \cellcolor[RGB]{192,255,192} -96.0\% & \cellcolor[RGB]{190,255,190} 98.5\% & \cellcolor[RGB]{237,255,237} 5.1\% & 38.5 & \cellcolor[RGB]{255,237,237} 5.0\% & \cellcolor[RGB]{192,255,192} -94.5\% & 112.5 & \cellcolor[RGB]{222,255,222} -35.4\% & \cellcolor[RGB]{190,255,190} -98.6\% & \cellcolor[RGB]{190,255,190} 99.0\% & \cellcolor[RGB]{237,255,237} 5.3\% \\
 & \cellcolor{lightgray}Fork & \cellcolor{lightgray}9.6 & \cellcolor{lightgray}\cellcolor[RGB]{230,255,230} -18.5\% & \cellcolor{lightgray}\cellcolor[RGB]{201,255,201} -76.6\% & \cellcolor{lightgray}14.7 & \cellcolor{lightgray}\cellcolor[RGB]{204,255,204} -70.2\% & \cellcolor{lightgray}\cellcolor[RGB]{195,255,195} -89.6\% & \cellcolor{lightgray}\cellcolor[RGB]{190,255,190} 99.0\% & \cellcolor{lightgray}\cellcolor[RGB]{232,255,232} 14.3\% & \cellcolor{lightgray}18.2 & \cellcolor{lightgray}\cellcolor[RGB]{239,255,239} -1.6\% & \cellcolor{lightgray}\cellcolor[RGB]{195,255,195} -88.4\% & \cellcolor{lightgray}30.2 & \cellcolor{lightgray}\cellcolor[RGB]{235,255,235} -8.3\% & \cellcolor{lightgray}\cellcolor[RGB]{192,255,192} -95.0\% & \cellcolor{lightgray}\cellcolor[RGB]{190,255,190} 99.0\% & \cellcolor{lightgray}\cellcolor[RGB]{232,255,232} 14.3\% \\
 & Group & 14.2 & \cellcolor[RGB]{238,255,238} -3.9\% & \cellcolor[RGB]{200,255,200} -78.3\% & 14.2 & \cellcolor[RGB]{255,237,237} 4.6\% & \cellcolor[RGB]{191,255,191} -96.8\% & \cellcolor[RGB]{201,255,201} 76.1\% & \cellcolor[RGB]{196,255,196} 86.8\% & 34.5 & \cellcolor[RGB]{238,255,238} -3.3\% & \cellcolor[RGB]{194,255,194} -90.4\% & 51.0 & \cellcolor[RGB]{238,255,238} -2.8\% & \cellcolor[RGB]{190,255,190} -99.1\% & \cellcolor[RGB]{202,255,202} 75.4\% & \cellcolor[RGB]{199,255,199} 80.5\% \\
 & \cellcolor{lightgray}Group Multiple & \cellcolor{lightgray}21.3 & \cellcolor{lightgray}\cellcolor[RGB]{239,255,239} -0.9\% & \cellcolor{lightgray}\cellcolor[RGB]{199,255,199} -80.1\% & \cellcolor{lightgray}35.6 & \cellcolor{lightgray}\cellcolor[RGB]{237,255,237} -5.6\% & \cellcolor{lightgray}\cellcolor[RGB]{190,255,190} -98.2\% & \cellcolor{lightgray}\cellcolor[RGB]{204,255,204} 71.2\% & \cellcolor{lightgray}\cellcolor[RGB]{193,255,193} 92.6\% & \cellcolor{lightgray}49.7 & \cellcolor{lightgray}\cellcolor[RGB]{255,239,239} 0.3\% & \cellcolor{lightgray}\cellcolor[RGB]{194,255,194} -90.7\% & \cellcolor{lightgray}98.6 & \cellcolor{lightgray}\cellcolor[RGB]{255,238,238} 2.8\% & \cellcolor{lightgray}\cellcolor[RGB]{190,255,190} -99.4\% & \cellcolor{lightgray}\cellcolor[RGB]{205,255,205} 68.1\% & \cellcolor{lightgray}\cellcolor[RGB]{196,255,196} 86.2\% \\
\end{tabular}}
\label{tab:workflow_results_1gbit}
\end{table*}
\endgroup

\subsubsection*{Real-world workflows}
The nf-core project~\cite{ewelsNfcoreFrameworkCommunitycurated2020a} is a community-curated collection of reusable workflows for Nextflow.
For our experiments, we use the three workflows with the most GitHub stars, all from the bioinformatics domain:
\begin{enumerate}
    \item \emph{RNA-Seq} is a workflow that analyses gene expression.
    We use data from a bladder cancer cells study~\cite{RNAseq_data}. %
    \item \emph{Sarek} is a workflow for variant calling, a genomic analysis identifying genetic variations. %
    We use data from a breast cancer study using CRISPR-Cas9 technology~\cite{sarek_data}. %
    \item \emph{Chip-Seq} is a workflow that allows us to study protein-DNA interactions and to understand different cellular processes. 
    We use data from an article that studies prostate cancers~\cite{chip_seq_data}. %
\end{enumerate}

As fourth real-life workflow outside Bioinformatics, we use \emph{Rangeland}~\cite{lehmannFORCENextflowScalable2021} from the remote sensing domain.
The Rangeland workflow analyses the 1984-2006 time series of Landsat satellite images to evaluate rangeland degradation in the Mediterranean region.
We run Rangeland on 304~GB of input data to study rangeland degradation for Crete, Greece.

\subsubsection*{Synthetic workflows}

WfCommons is a set of workflow tools, including WfChef~\cite{colemanWfChef}, which automatically generates realistic synthetic workflows synthesized from real traces, and WfBench~\cite{colemanwfBench}, which generates actual benchmark workflows for specific workflow engines.\looseness-1

We complemented our real-world workflows with synthetic workflow instances from WfChef to show our approach works well across different realistic workflow topologies.
Using synthetic workflows allows us to freely specify important workflow properties, like the number of tasks, task runtimes, input file sizes, and output file sizes.
We set the number of tasks to a limit of 200, adjusted the input to approximately 20~GB, and the output to approximately 150~GB.
Further, we set the CPU load such that the workflow is I/O bound.
Out of nine workflows provided by WfChef, we used only seven because two workflows use iterations in their DAG, which Nextflow does not support.
The remaining seven recipes include four recipes for bioinformatics workflow applications (i.e., 1000Genome, BLAST, BWA, and SoyKB), one from the agroecosystem domain (i.e., Cycles), one used in astronomy (i.e., Montage), and one specific to Seismology (i.e., Seismology).\looseness-1

\subsubsection*{Workflow patterns}
We generated five workflows that follow the main patterns occurring in workflows identified by \citeauthor{BharathiCharacterizationOfWorkflows}~\cite{BharathiCharacterizationOfWorkflows}, which are presented in Figure~\ref{fig:workflowPatterns}.

\begin{enumerate}[label=\alph*)]
    \item The ``All in One'' workflow comprises 100 tasks that write a file, and a subsequent task reads all the generated data.
    \item The ``Chain'' workflow involves 100 tasks followed each by another task reading the produced data.
    \item In the ``Fork'' workflow, one task has 100 successor tasks.
    \item The ``Group'' workflow has 100 tasks $A_i$, with $i = 1,\ldots,100$.
    These tasks are grouped by dividing their index $i$ by $3$, rounding down, and assigning them to the corresponding group: Group$_{floor(i/3)}$.
    \item Finally, the ``Group Multiple'' extends the ``Group'' workflow with a second grouping according to the rule: Group$_{floor(i/4)}$.
\end{enumerate}

\subsection{Execution Environment}

Our prototype is based on Nextflow in version 23.07.0-edge.
For our experiments, we set up a Kubernetes cluster with kubelet version 1.27.5 consisting of eight nodes.
Each node is equipped with an AMD EPYC 7282 16-core processor, 128~GB DDR4 memory, and two SATA III SSDs with 960~GB with approximately 537~MB/s read and 402~MB/s write in sequential mode.
From the two disks, every node contributes one SSD to a Ceph file system with a replica factor of two.
In addition, we provide a ninth node that exposes a 4TB PCIe 4.0 NVMe SSD via an NFS server. 
The nodes are physically connected via a 10~Gbit network, which we limit in our experiments to 1~Gbit or 2~Gbit using tcconfig to emulate a commodity cluster, in which network I/O is slower than storage access.

\subsection{Experimental Design}

We performed all experiments using the following design.

\emph{Median makespan measurements: }
For all runs in all experiments, we performed three repetitions and report the run with the median makespan - the time from the start of the first task to the end of the last task.

\emph{Baselines and WOW configuration: }
We compare three scheduling strategies: the Nextflow original scheduling (Orig), the Common Workflow Scheduler (CWS), and our approach (WOW).
Orig prioritizes tasks in a FIFO manner and assigns them in a RoundRobin fashion.
CWS prioritizes tasks by their rank and input size.
Both disregard data locations.
For the experiments, we set the COP constraints of our approach, $c^{node}$ and $c^{task}$, to one and two, respectively.

\textit{Experiments:}
We conduct three different experiments. 
\begin{enumerate}
    \item \textit{Execution behavior}: All 16 workflows were run with all three schedulers and two DFS configurations (Ceph and NFS) over a 1~Gbit network.
    \item \textit{Network dependence}: The execution behavior experiment was repeated with 2~Gbit bandwidth to assess network dependence. 
    This experiment used one real workflow ("Chip-Seq") and five patterns.
    \item \textit{Scalability efficiency}: Scalability was tested by reducing the number of nodes for workflows on a 1~Gbit network, comparing WOW to CWS.
\end{enumerate}

\section{Evaluation Results} 
This section presents the results from the three experiments we conducted.
All tables and plots can also be found in our GitHub repository\footref{foot::github}.\looseness=-1

\subsection{Execution Behavior}

Table~\ref{tab:workflow_results_1gbit} shows the results of running workflows with WOW, CWS, and the original Nextflow approach on eight nodes and a 1~Gbit network using either Ceph or NFS as the underlying storage system.

Taking the original Nextflow approach as a baseline, we see that the Common Workflow Scheduler improved the makespan for 11 out of 16 workflows.
In comparison, our WOW approach performs better than both competitors for all 16 workflows.

For real-world workflows, WOW was, for example, able to reduce the makespan by 18.3\% for ``RNA-Seq'' using Ceph as the storage platform and 53.3\% using NFS for data storage.
For the ``Rangeland'' workflow, the makespan reduction was 4.3\% with Ceph and 13.4\% with an NFS.

In the ``Chain'' pattern workflow, which is the optimal pattern for our approach, the makespan was reduced by 86.4\% when using Ceph as DFS and by 94.5\% when using NFS.
For the ``All in One'' pattern workflow, the makespan decreased by 49.3\% for Ceph and 60.1\% for NFS.

Also, in Table~\ref{tab:workflow_results_1gbit}, we compare the allocated CPU hours, calculated as the sum of all tasks' runtimes multiplied by their respective allocation of CPUs, regardless of their actual usage.
In this comparison, our WOW approach shows a reduction of up to 99.4\% for both synthetic workflows and workflow patterns and up to 60.8\% for real-world workflows.

\textit{Copy operations:}
For all workflows, at least 61.1\% of the tasks had all input data available already on the node on which the tasks are executed, so no files were copied, see Table~\ref{tab:workflow_results_1gbit}.
The numbers are almost identical for Ceph and NFS. 
For 16.0\% of all workflow tasks, COPs were created, but the tasks were started on a node where no COP for these tasks was needed.
This is always possible because multiple tasks can require the same data.
Hence, COPs for other tasks might already transfer the required data for a task to a node.
Similarly, sufficient resources can become available on the node where the data was generated after a COP was started.
For 85.5\% of the tasks, two or fewer COPs were required.

\textit{Data overhead:}
\begin{figure}
    \centering
       \includegraphics[width=0.99\columnwidth,trim={0mm 0mm 0mm 0mm},clip]{
       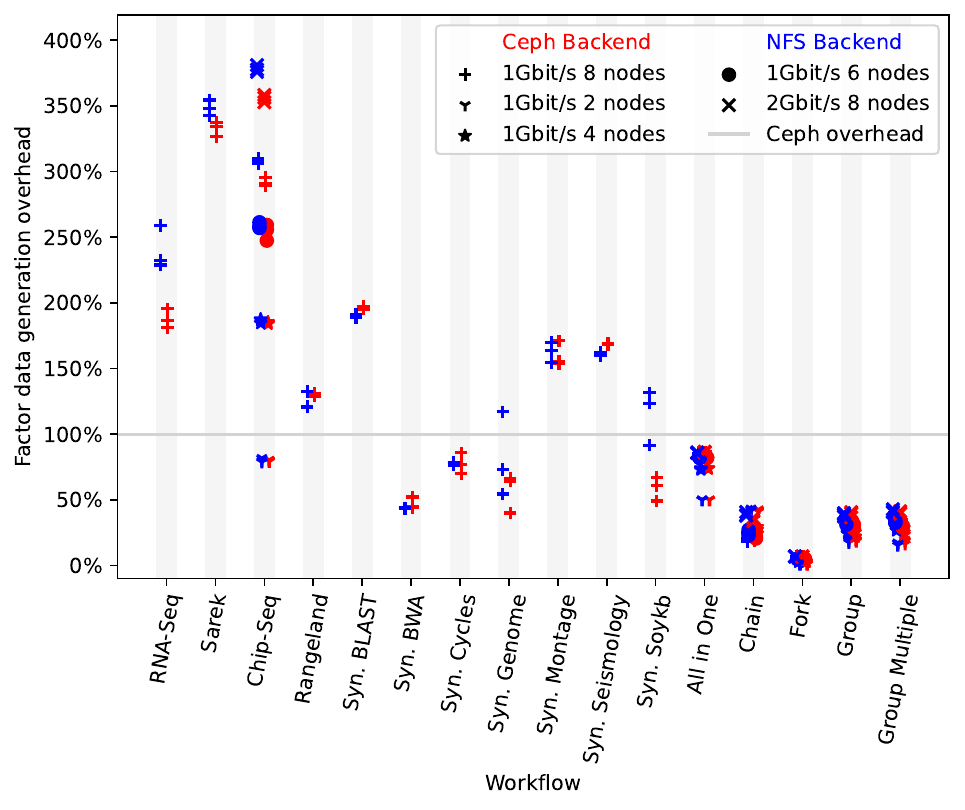}
    \caption{Data overhead of our approach. It is measured as the size of additional data replicas compared to the size of unique files. Red is the overhead when using Ceph as the backend for our approach, blue is the overhead when using NFS. The Ceph overhead is at 100\% since we use a replication level of two. NFS has no overhead as it introduces no replicas.}
    \label{fig:dataOverhead}
\end{figure}

We measure the data overhead WOW introduces and present it in Figure~\ref{fig:dataOverhead}.
For WOW, the data overhead translates to the amount of data copied through COPs, regardless of whether the copied files are used.
WOW's data overhead can be compared to the data overhead of Nextflow's original scheduling using NFS or Ceph: Using NFS as the file system means that the data is stored centrally and not replicated, so there is no overhead, while with Ceph, with a replication factor of two, as used in our experiments, the overhead is 100\%.
WOW's data overhead, in contrast, is variable and depends on the speculative copies performed.
We report the total overhead of data generated and, therefore, did not delete any replicas during our experiments.
However, these speculative temporary copies could be deleted as soon as the tasks requiring this data are finished.
For four out of seven synthetic workflows and all workflow patterns, the data overhead is smaller than for Ceph.
For real-world workflows, we see a larger overhead.
This is because tasks in these workflows perform more computations, and WOW uses the available time to perform speculative COPs.
It is also noteworthy that using NFS as the backend for WOW results in a larger overhead than using Ceph.
We assume this is because the first tasks read data from the DFS.
NFS is a single-point DFS, unlike Ceph, and the link to the NFS server is quickly overloaded.
Accordingly, task execution takes longer, giving the scheduler more time to prepare subsequent tasks.

\textit{Load distribution:}
We analyze the load distribution of our scheduling approach using the Gini coefficient, measuring equality in local storage usage and allocated CPU time across nodes. 
For real-world workflows, low Gini coefficients indicate balanced distributions: e.g., ``Rangeland'' (0.07) and ``Chip-Seq'' (0.01) for storage usage and ``Chip-Seq'' (0.00) for CPU time. 
The highest inequality is observed for ``Synthetic BWA'', where all tasks are executed on a single node since data transfer is longer than task execution.
Overall, we exhibit good balance, avoiding data and task hotspots.

\subsection{Network Dependence}
In our second experiment, we double the network bandwidth from 1~Gbit to 2~Gbit and show how this impacts makespan in Table~\ref{tab:workflow_results_2gbit}.
The idea is that methods that substantially reduce network traffic depend less on the available bandwidth and, therefore, see a lower decrease in makespan.

\textit{Execution behavior:}
The makespan improvements with increasing network bandwidth in the original Nextflow and CWS approaches are comparable, indicating that the network is a bottleneck for both approaches.
The ``All in One'' pattern has the highest reduction in makespan, with about 46.0\% for both competitors.
For the real workflow, ``Chip-Seq'', however, makespan is reduced by only 10\%.

For WOW, the makespan reduction is lower than for the competitors when the cluster network is changed from 1~Gbit to 2~Gbit.
This shows that WOW is less network-dependent and effectively reduces network bottlenecks, compared to the baselines.
For ``Chip-Seq'', makespan is not affected at all when using Ceph as the storage backend, indicating that we are transferring data before resources become available for tasks.
Thus no further improvement can be achieved.
For workflow patterns, this is different, as subsequent tasks often cannot start until all previous tasks have finished.
We do not see a comparable improvement for the ``Chain'' pattern, as all data is kept locally for this workflow.

The relative reduction in allocated CPU is similar to the relative decrease in makespan for the competitor strategies, showing that these workflows are predominantly I/O-bound.
In contrast, the change in network bandwidth does not impact the allocated CPU time for our approach, which is expected for patterns because the tasks read all data locally, and the local access speed remains the same across setups.
Furthermore, for ``Chip-Seq'', the allocated CPU time did not reduce even when we read the workflow input data from Ceph or NFS.
\begingroup
\setlength{\tabcolsep}{1.8mm}
\begin{table}
\centering
\caption{Relative change of the makespan of Chip-Seq workflow and workflow patterns when the network bandwidth is changed from 1Gbit to 2Gbit, for the three scheduling methods and two storage systems.
The highest makespan reduction for every method is marked in blue, and the lowest in red.\looseness-1}
\footnotesize
\resizebox{\columnwidth}{!}{
\begin{tabular}{l|r|r|r||r|r|r}
\multicolumn{1}{c|}{} & \multicolumn{3}{c||}{Ceph} & \multicolumn{3}{c}{NFS} \\
Workflow & Orig & CWS & WOW & Orig & CWS & WOW \\
\hline
All in One & \textcolor{blue}{-46.0\%} & \textcolor{blue}{-46.2\%} & \textcolor{blue}{-34.1\%} & -49.5\% & \textcolor{blue}{-49.6\%} & \textcolor{blue}{-33.1\%} \\
\cellcolor{lightgray}Chain & \cellcolor{lightgray}-27.5\% & \cellcolor{lightgray}-27.4\% & \cellcolor{lightgray}-2.0\% & \cellcolor{lightgray}\textcolor{blue}{-50.9\%} & \cellcolor{lightgray}-49.4\% & \cellcolor{lightgray}\textbf{\textcolor{red}{1.1\%}} \\
Chip-Seq & \textbf{\textcolor{red}{-7.9\%}} & \textbf{\textcolor{red}{-10.5\%}} & \textbf{\textcolor{red}{0.0\%}} & \textbf{\textcolor{red}{-31.5\%}} & \textbf{\textcolor{red}{-34.0\%}} & -9.6\% \\
\cellcolor{lightgray}Fork & \cellcolor{lightgray}-27.7\% & \cellcolor{lightgray}-28.7\% & \cellcolor{lightgray}-22.4\% & \cellcolor{lightgray}-47.5\% & \cellcolor{lightgray}-46.9\% & \cellcolor{lightgray}-16.8\% \\
Group & -34.9\% & -33.5\% & -23.0\% & -50.1\% & -47.1\% & -28.2\% \\
\cellcolor{lightgray}Group Multiple & \cellcolor{lightgray}-33.7\% & \cellcolor{lightgray}-37.0\% & \cellcolor{lightgray}-27.1\% & \cellcolor{lightgray}-48.8\% & \cellcolor{lightgray}-48.6\% & \cellcolor{lightgray}-32.7\% \\
\end{tabular}}
\label{tab:workflow_results_2gbit}
\end{table}
\endgroup

\textit{Copy operations:}
With the faster network, we see that more speculative COPs are executed.
Compared to 1~Gbit, we start a COP for 10\% more tasks, but the task ends up on a node where the required data is already in place.

\textit{Data overhead:}
From our results, we see that the data overhead of our approach increases as the network gets faster.
This is expected as it is less expensive to do speculative copies when more bandwidth is available, and WOW will hence be able to prepare high-priority tasks on more nodes.
The overhead increase is nearly identical for Ceph and NFS.
For the ``Chain'' pattern, we observe that about 82\% more data is copied.
In contrast, ``Fork'' and ``All in One'' do not have larger overheads, which, again, is expected since:
\begin{itemize}
    \item ``Fork'' copies the same file to all nodes and executes tasks everywhere.
    \item ``All in One'' makes two copies in parallel, starting as soon as the first one finishes; no third copy is instantiated, regardless of network speed.
\end{itemize}

\textit{Load distribution:}
As the network speed increases, the Gini coefficient for data distribution through WOW COPs becomes smaller.
In the ``Fork'' pattern, the Gini coefficient decreased by 0.08.
Equality is also better for ``Grouping'' and ``Grouping Multiple''.
The tasks-on-node-distribution remains similar as well.\looseness-1

Again, in the ``Fork'' pattern, the Gini coefficient decreases by 0.09.
For the allocated CPU hours per node, the change is also highest for the ``Fork'' pattern.

\subsection{Scalability Efficiency}

\begin{figure}
    \centering
    \begin{subfigure}{.5\columnwidth}
      \centering
        \includegraphics[height=3.65cm,trim={0mm 0mm 0mm 0mm},clip]{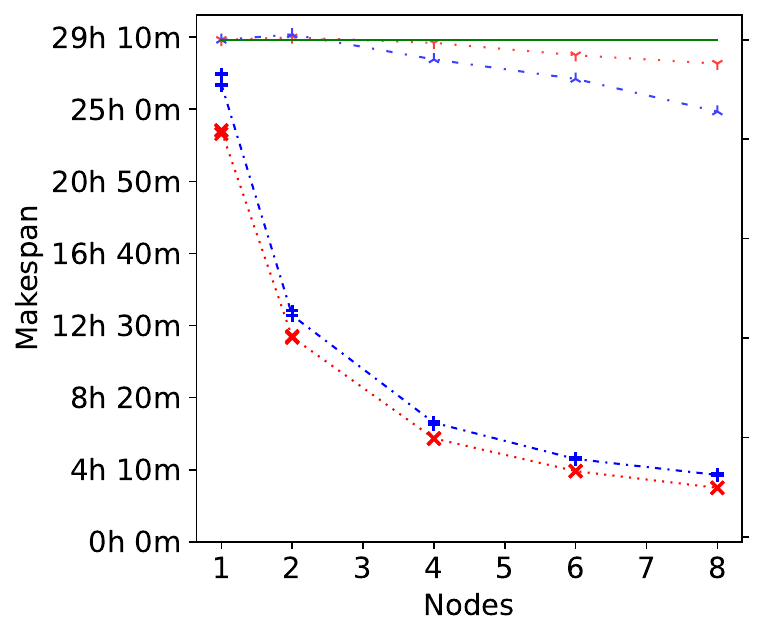}
      \caption{Chip-Seq Ceph}
	 \label{fig:makespan_efficiency_chipseq_ceph}
    \end{subfigure}%
    \hfill
    \begin{subfigure}{.5\columnwidth}
      \centering
         \includegraphics[height=3.65cm,trim={0mm 0mm 0mm 0mm},clip]{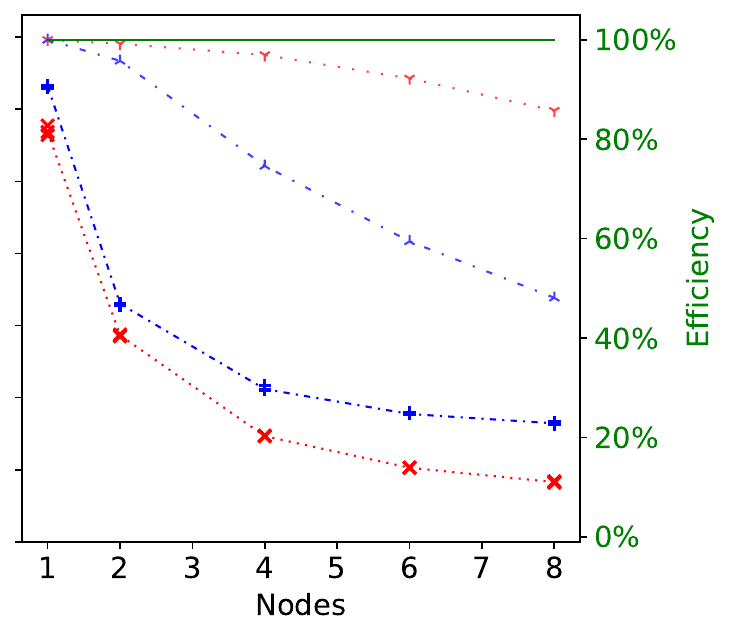}
      \caption{Chip-Seq NFS}
	 \label{fig:makespan_efficiency_chipseq_nfs}
    \end{subfigure}%

    \medskip
    
    \begin{subfigure}{.5\columnwidth}
      \centering
        \includegraphics[height=3.65cm,trim={0mm 0mm 0mm 0mm},clip]{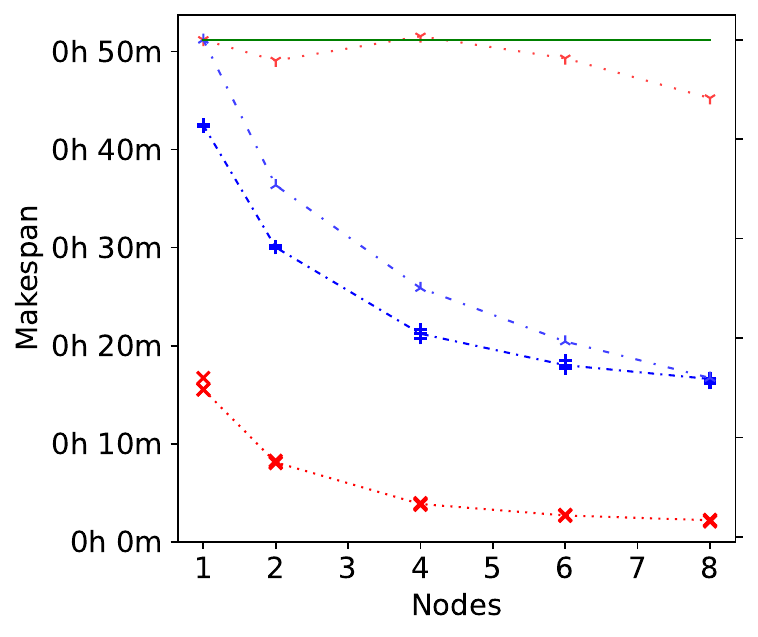}
      \caption{Chain Ceph}
	 \label{fig:makespan_efficiency_chain_ceph}
    \end{subfigure}%
    \hfill
    \begin{subfigure}{.5\columnwidth}
      \centering
        \includegraphics[height=3.65cm,trim={0mm 0mm 0mm 0mm},clip]{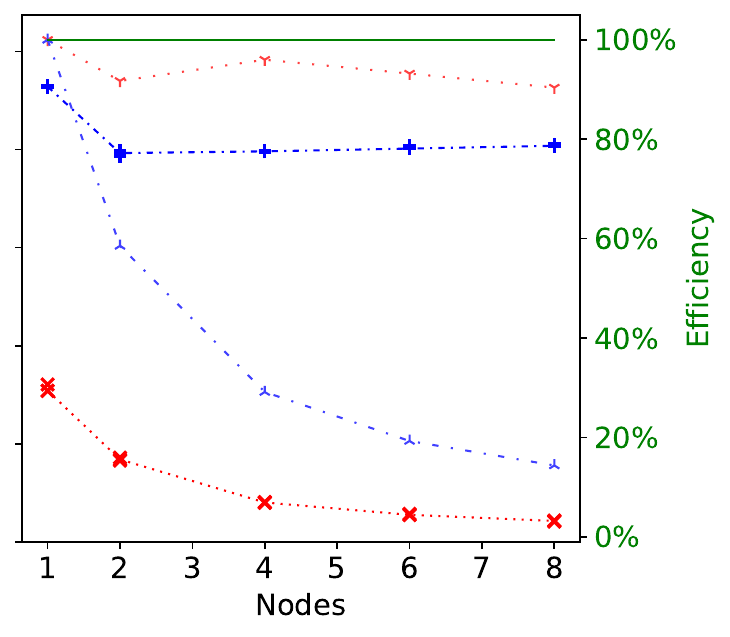}
      \caption{Chain NFS}
	 \label{fig:makespan_efficiency_chain_nfs}
    \end{subfigure}%
    
    \medskip
    
    \begin{subfigure}{.5\columnwidth}
      \centering
        \includegraphics[height=3.65cm,trim={0mm 0mm 0mm 0mm},clip]{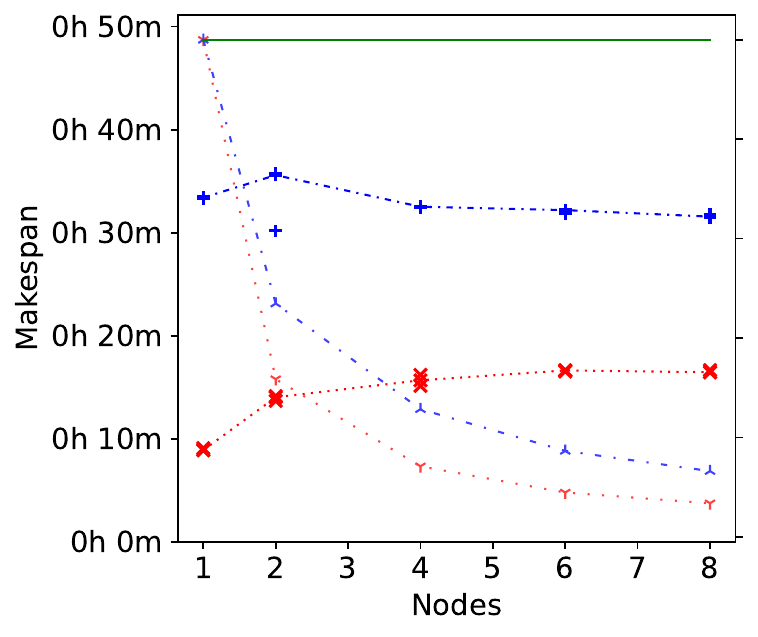}
      \caption{All Into One Ceph}
	 \label{fig:makespan_efficiency_allIntoOne_ceph}
    \end{subfigure}%
    \hfill
    \begin{subfigure}{.5\columnwidth}
      \centering
        \includegraphics[height=3.65cm,trim={0mm 0mm 0mm 0mm},clip]{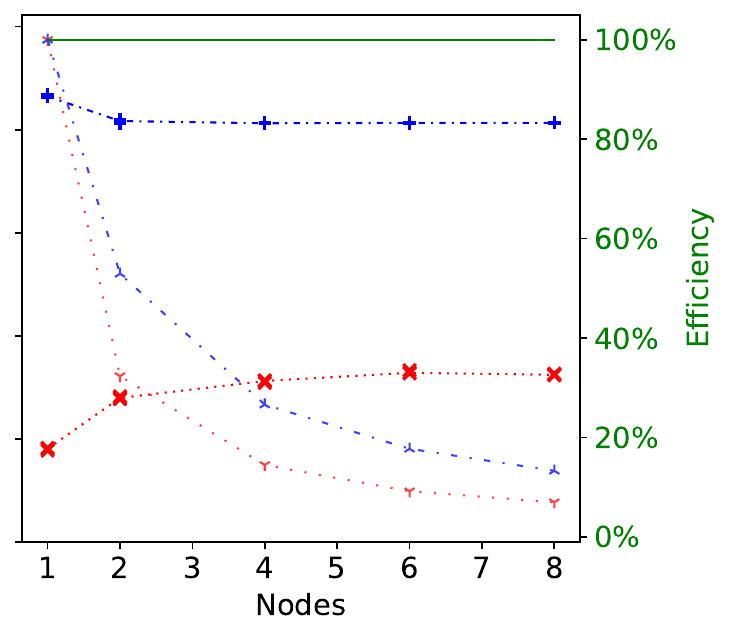}
      \caption{All In One NFS}
	 \label{fig:makespan_efficiency_allIntoOne_nfs}
    \end{subfigure}%

    \medskip
    
    \begin{subfigure}{0.8\columnwidth}
      \centering
      \includegraphics[width=\columnwidth,trim={17.9mm 116.0mm 3.3mm 4.3mm},clip]{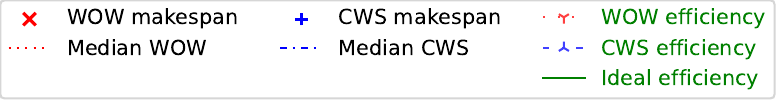}
    \end{subfigure}%
    \caption{Makespan and efficiency when scaling the number of nodes. An efficiency of 100\% indicates the single-node makespan divides exactly by the number of nodes used. WOW is presented in red and CWS scheduling in blue.}
    \label{fig::makespanEfficiency}
\end{figure}

We ran all the patterns and the ``Chip-Seq'' workflow on five different setups - one, two, four, six, and eight nodes.
Running on a single node is our baseline for efficiency. 
For WOW, there is no need to create a COP.
For the competitor strategies using the DFS for data exchange, the single node uses the DFS exclusively.
We expect increased overhead and reduced efficiency as we add more nodes.
To measure this, we define the efficiency for $n$ nodes as: $\displaystyle \text{efficiency(}n\text{)} = \frac{\text{makespan(}1\text{)}}{\text{makespan(}n\text{)} * n}$.
Accordingly, 100\% efficiency refers to an $n$ times reduction in makespan when running on $n$ nodes.

\textit{Execution behavior:}
Figure~\ref{fig::makespanEfficiency} shows the makespan and efficiency for the real-world workflow ``Chip-Seq'', the optimal pattern for WOW: ``Chain'', and the most challenging pattern: ``All in One''.
For ``Chip-Seq'' and ``Chain'', we see a high efficiency with WOW as we scale out, but efficiency decreases slightly for ``Chip-Seq'' and NFS because the input data is still read from NFS.
Using WOW instead of CWS, the scaling efficiency increases from 85.6\% to 96.2\% for ``Chip-Seq'' with Ceph and from 48.11\% to 85.7\% for NFS.
Again, increased efficiency while scaling out in a cluster is expected for NFS since it is single-point DFS, while Ceph is distributed across all nodes.
For ``Chain'', WOW remains at 90.3\% with Ceph and 88.3\% with NFS, while CWS drops to 32.0\% and 14.4\%, respectively.
Notably, the makespan does not decrease with CWS using NFS, showing that the connection to the NFS-server is at its maximum.\looseness-1

In contrast, for ``All in One'', the efficiency is 6.8\% for Ceph and 7.0\% for NFS when using WOW.
The CWS approach reaches 13.2\% and 13.3\%, respectively.

For all other workflow patterns, WOW consistently outperforms CWS with better efficiency and shorter makespan.

\textit{Copy operations:}
When we add more nodes, we see that a higher number of COPs are created, and a higher absolute number remains unused, as the tasks end up running on nodes where either other COPs copied all required data or where all required data is already present.
In comparison, the share of tasks that do not use any COP decreases from 21.6\% with two nodes to 16.0\% with eight nodes.

\textit{Data overhead:}
Especially for ``Chip-Seq'', a decrease in data overhead is visible as we use fewer nodes.
The workflow patterns also show a tendency toward less overhead with fewer nodes.
This is expected because tasks are less likely to need data from tasks running on another node, and if they do, the fraction of missing data becomes smaller.
Moreover, the number of possible replicas is limited by the number of nodes.
For example, with two nodes, we can never have more than 100\% overhead.\looseness-1

\textit{Load distribution:}
The fewer nodes we use, the better the Gini coefficient becomes for data distribution.
The performance of ``Chain'' and ``Chip-Seq'' did not improve because they already had a very low Gini coefficient.
The Gini coefficient for ``Fork'' is reduced from 0.14 for eight nodes to 0.02 for two nodes.
The distribution of tasks across nodes also becomes more equal as the number of tasks decreases, as does the distribution of CPU hours.
Again, the changes are particularly large for ``Fork'' and ``All in One'', which are the patterns where the workflow is inherently unbalanced, with one task in one stage and 100 tasks in another stage.

\section{Related Work}\label{sub:relatedWork}
While ample work has been done on workload-aware data placement, both from a file system and a middleware perspective, the specific combination of scientific workflow systems, resource managers, and distributed file systems offers unique challenges that, to the best of our knowledge, were not addressed comprehensively before.
In this section, we first describe other works targeting data placement in workflow systems, as these are most related to our work.
We then briefly discuss some other, less closely related topics and highlight the differences to our proposal.
We omit works presupposing a static scheduling, such as~\cite{sukhoroslovEfficientExecutionDataintensive2021,giampaDataAwareSchedulingStrategy2021,pietriSchedulingDataintensiveScientific2018}, as these are generally not applicable to our dynamic setting.

\subsection{Data Placement for Dynamic Scientific Workflows}
Several approaches targeting dynamic workflows assume all data to be stored in a central storage, from which they are fetched to local nodes prior to execution.
Local nodes only work as cache for these data.
For instance,~\cite{brykStorageawareAlgorithmsScheduling2016} proposes to schedule tasks to the node with the earliest expected finish time, also taking the necessary transfer time into account.
\cite{marozzoDataawareSchedulingStrategy2017}~describes a cloud-based data mining system that proceeds in a similar manner but makes scheduling decisions based on the proportion of locally available input data instead of modeling transfer times.
Such cache-based approaches face the problem that all intermediate results must first be written to the DFS, irrespective of which tasks require them next, and that placement decisions only take the next task into account.
Furthermore, as tasks are scheduled before the data is actually available, they can only start once the transfer is complete, causing unnecessary idle times.

\citeauthor{WangOptimizing2014} propose a task stealing scheduling approach to bring tasks to the node with the largest input share~\cite{WangOptimizing2014}.
However, the approach does not consider file transfer times.

Confuga is a batch execution and POSIX-compatible file system that also manages task dependencies and can thus be viewed as a workflow engine~\cite{donnellyConfugaScalableData2015}.
Similarly, TaskVine proposes an approach that focuses on task startup times~\cite{delgadoTaskVineManagingInCluster2023}.
Similar to WOW, both allow for parallel task scheduling and placement using asynchronous data transfer but rely on a simple FIFO scheduling queue and are implemented as a monolithic system, incompatible with real-life setups on existing file systems and resource managers.

\citeauthor{costaCaseWorkflowAwareStorage2015a} adapt parameters of the MosaStore for specific workflow patterns to test the opportunities of a workflow-aware file system~\cite{costaCaseWorkflowAwareStorage2015a}.
While this can only be optimized for one workflow pattern at a time, our prototype works with workflows combining several dataflow patterns.

ODDS~\cite{wangODDS2020} is a data-aware scheduler for scientific applications built upon the HDFS file system.
It monitors all data chunks in a system and schedules tasks so data movement is minimal.
However, ODDS only performs data-aware scheduling and does not optimize data placement itself.

The AMFS Shell~\cite{zhangParallelizingExecutionSequential2013} is a scripting approach to define the data dependencies between tasks in Bash scripts explicitly.
Therefore, it supports four movement strategies: multicast, gather, allgather, and scatter.
The AMFS system then copies the data to nodes using FTP, as we do.
AMFS is made for MPI applications and requires all data to fit into the memory.
In contrast, we do not store data in memory and thus do not require the data to fit.
Comparable to AMFS, \citeauthor{duroFlexibleDataAwareScheduling2016} extend the Swift scripting language for location-aware task scheduling~\cite{duroFlexibleDataAwareScheduling2016}.
Meanwhile, our approach does not need an explicit definition of data flow and works with dynamic workflows where tasks are submitted one after another.

FusionFS~\cite{zhaoFusionFSSupportingDataintensive2014} is a file system for workflows on HPC.
While it enables data-aware task scheduling, files are read remotely, when unavailable on a node.
While this works well for pipelines where files are read and written once, it is suboptimal for scientific workflows where files are accessed by different tasks.
Moreover, it again introduces the problem of unmanaged data transfer.\looseness-1

\subsection{Data Placement in Distributed Data Processing Systems} Distributed data processing systems, such as Apache Hadoop, Apache Spark, or Apache Flink, enable parallel processing of large amounts of data on distributed cluster infrastructures.
These systems offer predefined APIs that enable the structured definition of jobs within a framework of available operations, enhancing transparency and control over job execution in comparison to the more flexible but black-box nature of workflow systems.
Several works explored the interplay of data placement and operation scheduling in such systems.
For instance, CoHadoop~\cite{eltabakh2011cohadoop} is an extension for Hadoop~\cite{ShvachkoHadoop2010} that allows applications to control file storage locations, but does not perform placement decisions itself.
AdaptDB~\cite{Lu2017AdaptDBAP} is an adaptive storage manager for Spark that partitions datasets and refines data partitions during the execution.
This, however, requires the data to be structured in a known format, which is not the case for scientific workflows.

\section{Conclusion}
Keeping intermediate data locally on cluster nodes when executing scientific workflows can significantly reduce network bottlenecks.
Similarly, it is often more efficient to transfer required inputs directly to specific nodes than to exchange such data via networked storage.
Based on these observations, we developed the WOW approach, where task scheduling and data placement go hand in hand, task input data is copied speculatively in preparation for the actual task execution, and the overall network traffic is considerably reduced.

The results of our empirical evaluation of a WOW prototype implementation for Nextflow and Kubernetes are promising and show a runtime reduction between 49.3\% and 94.5\% for frequently used workflow patterns.
Meanwhile, for real-world workflows, WOW reduced the runtime by up to 53.2\%.
Also, by separating copy operations and task executions in time, WOW reduces the allocated CPU time by up to 99.4\% for frequently used workflow patterns and up to 60.8\% for real-world workflows. 

Our evaluation results further indicate that WOW distributes both data and computation fairly across clusters, with Gini coefficients for load distributions close to zero on average over all workflows evaluated.
Moreover, we show that our approach can exhibit significantly better scaling behavior than strategies that keep data in a DFS.
In addition, our WOW approach is less dependent on the available network bandwidth and the available DFS, which often cannot be configured by individual users in multi-tenant clusters.

WOW is currently limited to homogeneous clusters where all nodes execute tasks at similar speeds.
While we focus on single clusters in this paper, applying our approach to workflows running across multiple clusters is an idea for future work.
Moreover, we plan to improve fault tolerance by strategically placing additional replicas of all intermediate files.
Nevertheless, our experimental results show that WOW already significantly reduces the makespan, CPU usage, and network bottlenecks of a variety of real-world and synthetic workflows in a single cluster, for two relevant storage systems and compared to state-of-the-art baselines.

\section*{Acknowledgment}
This work was funded by the German Research Council (DFG) as part of the  CRC 1404: ``FONDA: Foundations of Workflows for Large-Scale Scientific Data Analysis.''
\putbib
\end{bibunit}

\clearpage

\begin{bibunit}
\appendix
\subsection{Abstract}

In this artifact, we describe how to build the WOW Scheduler for Kubernetes, which is based on the Common Workflow Scheduler~\cite{lehmannHowWorkflowEngines2023}, and how to build and use the adapted Nextflow\cite{ditommasoNextflowEnablesReproducible2017} version, both of which are proposed in the paper: ``WOW: Workflow-Aware Data Movement and Task Scheduling for Dynamic Scientific Workflows''.
Moreover, we describe the experimental setup and how to prepare and get the input data of the 16 presented workflows and run them with the Original, CWS, and WOW strategies.

\subsection{Description}

\subsubsection{Check-List (Artifact Meta Information)}

{\small
\begin{itemize}
  \item {\bf Algorithm: } The WOW strategy is implemented into the Common Workflow Scheduler for Kubernetes.
  \item {\bf Program: } WOW for Kubernetes~\cite{wow_Scheduler} and Nextflow with CWS and WOW extension for Kubernetes~\cite{wow_Nextflow}.
  \item {\bf Compilation: } Nextflow with CWS and WOW extension for Kubernetes and our Scheduler for Kubernetes as Docker image.
  \item {\bf Data set: } Scripts to download the input data and execute the experiments~\cite{wow_ExperimentSetup}. Traces and logs of all 648 workflow executions~\cite{wow_Results}.
  \item {\bf Run-time environment: } Kubernetes cluster in version \emph{1.27.5}.
  \item {\bf Hardware: } All nodes were x86-64 machines.
  \item {\bf Runtime state: } The cluster was exclusively used for the experiments.
  \item {\bf Execution: } Bash scripts to manage the experiment, YAML files to set up Kubernetes, Dockerfiles, and scripts to build Nextflow and the Scheduler.
  \item {\bf Output: } Workflow traces and logs of all 648 workflow executions that are processed to generate the tables and plots in the paper.
  \item {\bf Experiment workflow: } Download the inputs and run the execution file.
  \item {\bf Experiment customization: } The experiments can be executed with different workflows, other data sets, or other shared storage.
  \item {\bf Publicly available?: } Yes, all code and experimental results are hosted by us. Input data is obtained from public sources.
\end{itemize}
}

\subsubsection{How Software Can Be Obtained}
We created two major software artifacts: the scheduler and our adapted Nextflow. 

The Scheduler for the Kubernetes implementation can be cloned from GitHub~\href{https://github.com/WOW-WorkflowScheduler/KubernetesScheduler}{https://github.com/WOW-WorkflowScheduler/nextflow}.
Moreover, we provide an already-built Docker image (commonworkflowscheduler/kubernetesscheduler:v1.0) starting the Kubenetes Scheduler service.

The adapted Nextflow version can be cloned from GitHub~\href{https://github.com/WOW-WorkflowScheduler/nextflow}{https://github.com/WOW-WorkflowScheduler/nextflow}.
Again, a prebuilt Docker image is available on DockerHub (commonworkflowscheduler/nextflow-wow:v1.0).

\subsubsection{Hardware Dependencies}
All artifacts are tested and prebuilt with x86-64 machines running Ubuntu \emph{22.04 LTS}. 

\subsubsection{Software Dependencies}
Building the software requires installing Docker and Java OpenJDK 19.
The Scheduler is based on CWS and uses Maven as a build system. 
Maven only needs to be installed if CWS is not built using Docker.
Nextflow uses Gradle as a build system and make - Gradle does not need to be installed, but the JDK needs to be in a matching version.
Moreover, a Kubernetes Cluster with enough local storage on each node and a shared file system is required.
The required amount of local storage depends on the number of nodes and the amount of generated intermediate data.
For our experiments, we used Ceph and NFS as shared filesystems. 
However, any read-write-many filesystem supported by Kubernetes and Nextflow will work but will change the results.
We need kubectl installed to communicate with the cluster and launch the experiment.

\subsubsection{Datasets}
We use the public datasets for real-world workflows, and for synthetic workflows, we generate synthetic data.
The input data is always prefetched to avoid affecting experiment runtimes with download times.
The download is performed by executing the \emph{setup-inputs.sh} file as described in~\href{https://github.com/WOW-WorkflowScheduler/Experiments}{https://github.com/WOW-WorkflowScheduler/Experiments}.
Moreover, we provide a configuration file for each workflow.

\subsection{Installation}

\subsubsection{Build Nextflow}
Run the following instructions to build and publish the Nextflow with CWS and WOW Docker image.
\begin{lstlisting}[language=bash]
# Set JAVA_HOME to version 19
$ cd <nextflow root directory>
$ make compile
$ make pack
$ make install
$ make dockerPack
# login to Docker
$ docker tag nextflow/nextflow:23.07.0-edge <your docker account>/nextflow:<version>
\end{lstlisting}

\subsubsection{Build Common Workflow Scheduler with WOW}
To build the CWS with WOW, run the following commands.
\begin{lstlisting}[language=bash]
$ cd <Kubernetes Scheduler root directory>
$ docker build -t wow-scheduler .
$ docker tag wow-scheduler <your docker account>/wow-scheduler:<version>
\end{lstlisting}

\subsubsection{Prepare the Cluster}
To prepare the cluster, clone the following project: \href{https://github.com/WOW-WorkflowScheduler/Experiments}{https://github.com/WOW-WorkflowScheduler/Experiments}.

To prepare the execution of the experiments:
First, set your NFS server's IP and username in \emph{experiment/nfsConnection.txt} and create a file \emph{experiment/nfsPassword.txt} and insert your NFS password; this will be used to adjust the network speed of the server.
Then, also set the NFS server's IP in \emph{experiment/nfs.yaml}. 
Next, change the Kubernetes namespace in \emph{experiment/namespace.txt} and in \emph{experiment/accounts.yaml}, \emph{experiment/nfsClaim.yaml}, \emph{setup/dowload-pod.yaml}, maybe adjust your storage classes.
Further, label your Kubernetes nodes and adjust \emph{experiment/nextflow\_usedby.config} and \emph{setup/download-pod.yaml} accordingly.
To label the nodes, run the following:
\begin{lstlisting}[language=bash]
# Nodes executing tasks
kubectl label nodes <Node 1> <Node 2> usedby=<your name>
# Node to run the scheduler and Nextflow
kubectl label nodes <Node 0> management=true
\end{lstlisting}

Configure the NFS Server for Kubernetes by applying the following commands:
\begin{lstlisting}[language=bash]
$ kubectl apply -f experiment/nfs.yaml
$ kubectl apply -f experiment/nfsClaim.yaml
\end{lstlisting}

In the \emph{setup} directory, you find the download-pod.
Make sure an NFS server or another PVC exists where the data will be stored.
The PVC should match the \emph{claimName} in the download-pod specification.
Next, start the download pod:
\begin{lstlisting}[language=bash]
$ kubectl create -f download-pod.yaml
\end{lstlisting}

To download the data, run the setup-input script.
\begin{lstlisting}[language=bash]
$ bash setup-inputs.sh <workflow-name>
\end{lstlisting}

\subsubsection{Running the Experiments}

To execute the experiments, use the runExperimentFromRemote script in the experiment directory.
Launch experiments with:
\begin{lstlisting}[language=bash]
$ bash runExperimentFromRemote.sh <strategy> <network speed>
\end{lstlisting}

where strategy is nfs, ceph, orig-nfs, or orig-ceph. The network speed is an integer.
The script will do the following:
\begin{enumerate}
    \item Adjust the network connection between all nodes
    \item Run all workflows and do for all:
    \item Load all required images into Kubernetes' cache on all nodes
    \item Run the workflow for the provided strategy
    \item Collect all results
\end{enumerate}

\subsection{Evaluation and Expected Result}

All our measurements can be found in the \emph{result} folder of the results and evaluation GitHub: \href{https://github.com/WOW-WorkflowScheduler/ResultsAndEvaluation}{https://github.com/WOW-WorkflowScheduler/ResultsAndEvaluation}.
For each run, we collected the DAG, the amount of data stored at the node and at the PVC, Nextflow's logs, the config used, the report generated by Nextflow, the scheduler's log, a timeline.html, Nextflow's trace, and for runs with WOW the copy tasks created.

Besides the measurements, we provide the Jupyter script (\emph{Evaluation.ipynb}) that generated the plots and tables in the paper.

\subsection{Experiment Customization}
The datasets can be changed, or other workflows can be included.
Therefore, the data needs to be downloaded, and the workflow needs to be added in the runExperiments.sh.

\subsection{Future Development}
We merged the WOW strategy into the Common Workflow Scheduler and extended the nf-cws\footnote{\url{https://github.com/CommonWorkflowScheduler/nf-cws}} Plugin to replace our separate Nextflow fork.
Future developments will be based on these two projects.

\putbib
\end{bibunit}


\begin{thebibliography}{33}
\providecommand{\natexlab}[1]{#1}
\providecommand{\url}[1]{#1}
\csname url@samestyle\endcsname
\providecommand{\newblock}{\relax}
\providecommand{\bibinfo}[2]{#2}
\providecommand{\BIBentrySTDinterwordspacing}{\spaceskip=0pt\relax}
\providecommand{\BIBentryALTinterwordstretchfactor}{4}
\providecommand{\BIBentryALTinterwordspacing}{\spaceskip=\fontdimen2\font plus
\BIBentryALTinterwordstretchfactor\fontdimen3\font minus \fontdimen4\font\relax}
\providecommand{\BIBforeignlanguage}[2]{{%
\expandafter\ifx\csname l@#1\endcsname\relax
\typeout{** WARNING: IEEEtranN.bst: No hyphenation pattern has been}%
\typeout{** loaded for the language `#1'. Using the pattern for}%
\typeout{** the default language instead.}%
\else
\language=\csname l@#1\endcsname
\fi
#2}}
\providecommand{\BIBdecl}{\relax}
\BIBdecl

\bibitem[Singh et~al.(2019)Singh, Graves, Anantharaj, and Sukumar]{singh2019evaluating}
R.~Singh, J.~A. Graves, V.~Anantharaj, and S.~R. Sukumar, ``Evaluating {{Scientific Workflow Engines}} for {{Data}} and {{Compute Intensive Discoveries}},'' in \emph{2019 {{IEEE Big Data}}}, 2019.

\bibitem[Costa et~al.(2015)Costa, Yang, Vairavanathan, Barros, Maheshwari, Fedak, Katz, Wilde, Ripeanu, and {Al-Kiswany}]{costaCaseWorkflowAwareStorage2015a}
L.~B. Costa, H.~Yang, E.~Vairavanathan, A.~Barros, K.~Maheshwari, G.~Fedak, D.~Katz, M.~Wilde, M.~Ripeanu, and S.~{Al-Kiswany}, ``The {{Case}} for {{Workflow-Aware Storage}}:{{An Opportunity Study}},'' \emph{Journal of Grid Computing}, vol.~13, no.~1, 2015.

\bibitem[N'Takpé et~al.(2022)N'Takpé, Edgard~Gnimassoun, Oumtanaga, and Suter]{ntakpeDataawareSimulationdrivenPlanning2022}
T.~N'Takpé, J.~Edgard~Gnimassoun, S.~Oumtanaga, and F.~Suter, ``Data‐aware and simulation‐driven planning of scientific workflows on {{IaaS}} clouds,'' \emph{Concurrency and Computation: Practice and Experience}, vol.~34, no.~14, 2022.

\bibitem[Sukhoroslov(2021)]{sukhoroslovEfficientExecutionDataintensive2021}
O.~Sukhoroslov, ``Toward efficient execution of data-intensive workflows,'' \emph{The Journal of Supercomputing}, vol.~77, no.~8, 2021.

\bibitem[Subedi et~al.(2018)Subedi, Davis, Duan, Klasky, Kolla, and Parashar]{subediStackerAutonomicData2018}
P.~Subedi, P.~Davis, S.~Duan, S.~Klasky, H.~Kolla, and M.~Parashar, ``Stacker: {{An Autonomic Data Movement Engine}} for {{Extreme-Scale Data Staging-Based In-Situ Workflows}},'' in \emph{{{SC18}}: {{Int. Conf.}} for {{High Performance Computing}}, {{Networking}}, {{Storage}} and {{Analysis}}}, 2018.

\bibitem[Cima et~al.(2018)Cima, Böhm, Martinovič, Dvorský, Janurová, Aa, Ashby, and Chupakhin]{cimaHyperLoomPlatformDefining2018}
V.~Cima, S.~Böhm, J.~Martinovič, J.~Dvorský, K.~Janurová, T.~V. Aa, T.~J. Ashby, and V.~Chupakhin, ``{{HyperLoom}}: {{A Platform}} for {{Defining}} and {{Executing Scientific Pipelines}} in {{Distributed Environments}},'' in \emph{Proc. of the 9th {{Workshop}} and 7th {{Workshop}} on {{Parallel Programming}} and {{RunTime Management Techniques}} for {{Manycore Architectures}} and {{Design Tools}} and {{Architectures}} for {{Multicore Embedded Computing Platforms}}}, 2018.

\bibitem[{Sly-Delgado} et~al.(2023){Sly-Delgado}, Phung, Thomas, Simonetti, Hennessee, Tovar, and Thain]{delgadoTaskVineManagingInCluster2023}
B.~{Sly-Delgado}, T.~S. Phung, C.~Thomas, D.~Simonetti, A.~Hennessee, B.~Tovar, and D.~Thain, ``{{TaskVine}}: {{Managing In-Cluster Storage}} for {{High-Throughput Data Intensive Workflows}},'' in \emph{Proc. of the {{SC}} '23 {{Workshops}} of {{The Int. Conf.}} on {{High Performance Computing}}, {{Network}}, {{Storage}}, and {{Analysis}}}, 2023.

\bibitem[Topcuoglu et~al.(2002)Topcuoglu, Hariri, and {Min-You Wu}]{topcuogluPerformanceeffectiveLowcomplexityTask2002}
H.~Topcuoglu, S.~Hariri, and {Min-You Wu}, ``Performance-effective and low-complexity task scheduling for heterogeneous computing,'' \emph{IEEE TPDS}, vol.~13, no.~3, 2002.

\bibitem[Pietri and Sakellariou(2018)]{pietriSchedulingDataintensiveScientific2018}
I.~Pietri and R.~Sakellariou, ``Scheduling data-intensive scientific workflows with reduced communication,'' in \emph{Proc. of the 30th {{Int. Conf.}} on {{Scientific}} and {{Statistical Database Management}}}, 2018.

\bibitem[Liu et~al.(2020)Liu, Lu, and Che]{liu2020survey}
J.~Liu, S.~Lu, and D.~Che, ``A {{Survey}} of {{Modern Scientific Workflow Scheduling Algorithms}} and {{Systems}} in the {{Era}} of {{Big Data}},'' in \emph{2020 {{IEEE SCC}}}, 2020.

\bibitem[Liu et~al.(2015)Liu, Pacitti, Valduriez, and Mattoso]{liuSurveyDataIntensiveScientific2015a}
J.~Liu, E.~Pacitti, P.~Valduriez, and M.~Mattoso, ``A {{Survey}} of {{Data-Intensive Scientific Workflow Management}},'' \emph{Journal of Grid Computing}, vol.~13, no.~4, 2015.

\bibitem[Witt et~al.(2019)Witt, Van~Santen, and Leser]{wittLearning2019}
C.~Witt, J.~Van~Santen, and U.~Leser, ``Learning {{Low-Wastage Memory Allocations}} for {{Scientific Workflows}} at {{IceCube}},'' in \emph{2019 {{Int. Conf.}} on {{High Performance Computing}} \& {{Simulation}} ({{HPCS}})}, 2019.

\bibitem[Donnelly et~al.(2015)Donnelly, Hazekamp, and Thain]{donnellyConfugaScalableData2015}
P.~Donnelly, N.~Hazekamp, and D.~Thain, ``Confuga: {{Scalable Data Intensive Computing}} for {{POSIX Workflows}},'' in \emph{{Proc. CCGrid 2015}}, 2015.

\bibitem[Lehmann et~al.(2023)Lehmann, Bader, Tschirpke, Thamsen, and Leser]{lehmannHowWorkflowEngines2023}
F.~Lehmann, J.~Bader, F.~Tschirpke, L.~Thamsen, and U.~Leser, ``How {{Workflow Engines Should Talk}} to {{Resource Managers}}: {{A Proposal}} for a {{Common Workflow Scheduling Interface}},'' in \emph{{Proc. CCGrid 2023}}, 2023.

\bibitem[Ewels et~al.(2020)Ewels, Peltzer, Fillinger, Patel, Alneberg, Wilm, Garcia, Di~Tommaso, and Nahnsen]{ewelsNfcoreFrameworkCommunitycurated2020a}
P.~A. Ewels, A.~Peltzer, S.~Fillinger, H.~Patel, J.~Alneberg, A.~Wilm, M.~U. Garcia, P.~Di~Tommaso, and S.~Nahnsen, ``The nf-core framework for community-curated bioinformatics pipelines,'' \emph{Nature Biotechnology}, vol.~38, no.~3, 2020.

\bibitem[Green et~al.(2021)Green, Osterhout, Klova, Merkwirth, McDonnell, Zavareh, Fuchs, Kamal, and Jakobsen]{RNAseq_data}
J.~L. Green, R.~E. Osterhout, A.~L. Klova, C.~Merkwirth, S.~R. McDonnell, R.~B. Zavareh, B.~C. Fuchs, A.~Kamal, and J.~S. Jakobsen, ``Molecular characterization of type {{I IFN-induced}} cytotoxicity in bladder cancer cells reveals biomarkers of resistance,'' \emph{Molecular Therapy - Oncolytics}, vol.~23, 2021.

\bibitem[Harrod et~al.(2022)Harrod, Lai, Goldsbrough, Simmons, Oppermans, Santos, Győrffy, Allsopp, Toghill, Balachandran, Lawson, Morrow, Surakala, Carnevalli, Zhang, Guttery, Shaw, Coombes, Buluwela, and Ali]{sarek_data}
A.~Harrod, C.-F. Lai, I.~Goldsbrough, G.~M. Simmons, N.~Oppermans, D.~B. Santos, B.~Győrffy, R.~C. Allsopp, B.~J. Toghill, K.~Balachandran, M.~Lawson, C.~J. Morrow, M.~Surakala, L.~S. Carnevalli, P.~Zhang, D.~S. Guttery, J.~A. Shaw, R.~C. Coombes, L.~Buluwela, and S.~Ali, ``Genome engineering for estrogen receptor mutations reveals differential responses to anti-estrogens and new prognostic gene signatures for breast cancer,'' \emph{Oncogene}, vol.~41, no.~44, 2022.

\bibitem[Baumgart et~al.(2020)Baumgart, Nevedomskaya, Lesche, Newman, Mumberg, and Haendler]{chip_seq_data}
S.~J. Baumgart, E.~Nevedomskaya, R.~Lesche, R.~Newman, D.~Mumberg, and B.~Haendler, ``Darolutamide antagonizes androgen signaling by blocking enhancer and super‐enhancer activation,'' \emph{Molecular Oncology}, vol.~14, no.~9, 2020.

\bibitem[Lehmann et~al.(2021)Lehmann, Frantz, Becker, Leser, and Hostert]{lehmannFORCENextflowScalable2021}
F.~Lehmann, D.~Frantz, S.~Becker, U.~Leser, and P.~Hostert, ``{{FORCE}} on {{Nextflow}}: {{Scalable Analysis}} of {{Earth Observation}} data on {{Commodity Clusters}},'' in \emph{Proc. of the {{CIKM}} 2021 {{Workshops}}}, ser. {{CEUR}} Workshop Proc., vol. 3052, 2021.

\bibitem[Coleman et~al.(2021)Coleman, Casanova, and Da~Silva]{colemanWfChef}
T.~Coleman, H.~Casanova, and R.~F. Da~Silva, ``{{WfChef}}: {{Automated Generation}} of {{Accurate Scientific Workflow Generators}},'' in \emph{2021 {{IEEE}} 17th {{Int. Conf.}} on {{eScience}} ({{eScience}})}, 2021.

\bibitem[Coleman et~al.(2022)Coleman, Casanova, Maheshwari, Pottier, Wilkinson, Wozniak, Suter, Shankar, and Da~Silva]{colemanwfBench}
T.~Coleman, H.~Casanova, K.~Maheshwari, L.~Pottier, S.~R. Wilkinson, J.~Wozniak, F.~Suter, M.~Shankar, and R.~F. Da~Silva, ``{{WfBench}}: {{Automated Generation}} of {{Scientific Workflow Benchmarks}},'' in \emph{2022 {{IEEE}}/{{ACM Int. Workshop}} on {{Performance Modeling}}, {{Benchmarking}} and {{Simulation}} of {{High Performance Computer Systems}} ({{PMBS}})}, 2022.

\bibitem[Bharathi et~al.(2008)Bharathi, Chervenak, Deelman, Mehta, Su, and Vahi]{BharathiCharacterizationOfWorkflows}
S.~Bharathi, A.~Chervenak, E.~Deelman, G.~Mehta, M.-H. Su, and K.~Vahi, ``Characterization of scientific workflows,'' in \emph{2008 {{Third Workshop}} on {{Workflows}} in {{Support}} of {{Large-Scale Science}}}, 2008.

\bibitem[Giampa et~al.(2021)Giampa, Belcastro, Marozzo, Talia, and Trunfio]{giampaDataAwareSchedulingStrategy2021}
S.~Giampa, L.~Belcastro, F.~Marozzo, D.~Talia, and P.~Trunfio, ``A {{Data-Aware Scheduling Strategy}} for {{Executing Large-Scale Distributed Workflows}},'' \emph{IEEE Access}, vol.~9, 2021.

\bibitem[Bryk et~al.(2016)Bryk, Malawski, Juve, and Deelman]{brykStorageawareAlgorithmsScheduling2016}
P.~Bryk, M.~Malawski, G.~Juve, and E.~Deelman, ``Storage-aware {{Algorithms}} for {{Scheduling}} of {{Workflow Ensembles}} in {{Clouds}},'' \emph{Journal of Grid Computing}, vol.~14, no.~2, 2016.

\bibitem[Marozzo et~al.(2017)Marozzo, Rodrigo~Duro, Garcia~Blas, Carretero, Talia, and Trunfio]{marozzoDataawareSchedulingStrategy2017}
F.~Marozzo, F.~Rodrigo~Duro, J.~Garcia~Blas, J.~Carretero, D.~Talia, and P.~Trunfio, ``A data‐aware scheduling strategy for workflow execution in clouds,'' \emph{Concurrency and Computation: Practice and Experience}, vol.~29, no.~24, 2017.

\bibitem[Wang et~al.(2014)Wang, Zhou, Li, Zhao, Lang, and Raicu]{WangOptimizing2014}
K.~Wang, X.~Zhou, T.~Li, D.~Zhao, M.~Lang, and I.~Raicu, ``Optimizing load balancing and data-locality with data-aware scheduling,'' in \emph{2014 {{IEEE Big Data}}}, 2014.

\bibitem[Wang et~al.(2020)Wang, Han, Yin, Zhou, and Jiang]{wangODDS2020}
J.~Wang, D.~Han, J.~Yin, X.~Zhou, and C.~Jiang, ``{{ODDS}}: {{Optimizing Data-Locality Access}} for {{Scientific Data Analysis}},'' \emph{IEEE Transactions on Cloud Computing}, vol.~8, no.~1, 2020.

\bibitem[Zhang et~al.(2013)Zhang, Katz, Armstrong, Wozniak, and Foster]{zhangParallelizingExecutionSequential2013}
Z.~Zhang, D.~S. Katz, T.~G. Armstrong, J.~M. Wozniak, and I.~Foster, ``Parallelizing the execution of sequential scripts,'' in \emph{Proc. of the {{Int. Conf.}} on {{High Performance Computing}}, {{Networking}}, {{Storage}} and {{Analysis}}}, 2013.

\bibitem[Duro et~al.(2016)Duro, Blas, Isaila, Wozniak, Carretero, and Ross]{duroFlexibleDataAwareScheduling2016}
F.~R. Duro, J.~G. Blas, F.~Isaila, J.~M. Wozniak, J.~Carretero, and R.~Ross, ``Flexible {{Data-Aware Scheduling}} for {{Workflows}} over an {{In-memory Object Store}},'' in \emph{{Proc. CCGrid 2016}}, 2016.

\bibitem[Zhao et~al.(2014)Zhao, Zhang, Zhou, Li, Wang, Kimpe, Carns, Ross, and Raicu]{zhaoFusionFSSupportingDataintensive2014}
D.~Zhao, Z.~Zhang, X.~Zhou, T.~Li, K.~Wang, D.~Kimpe, P.~Carns, R.~Ross, and I.~Raicu, ``{{FusionFS}}: {{Toward}} supporting data-intensive scientific applications on extreme-scale high-performance computing systems,'' in \emph{2014 {{IEEE Big Data}}}, 2014.

\bibitem[Eltabakh et~al.(2011)Eltabakh, Tian, Özcan, Gemulla, Krettek, and McPherson]{eltabakh2011cohadoop}
M.~Y. Eltabakh, Y.~Tian, F.~Özcan, R.~Gemulla, A.~Krettek, and J.~McPherson, ``{{CoHadoop}}: Flexible data placement and its exploitation in {{Hadoop}},'' \emph{Proc. of the VLDB Endowment}, vol.~4, no.~9, 2011.

\bibitem[Shvachko et~al.(2010)Shvachko, Kuang, Radia, and Chansler]{ShvachkoHadoop2010}
K.~Shvachko, H.~Kuang, S.~Radia, and R.~Chansler, ``The {{Hadoop Distributed File System}},'' in \emph{2010 {{IEEE}} 26th {{Symposium}} on {{Mass Storage Systems}} and {{Technologies}} ({{MSST}})}, 2010.

\bibitem[Lu et~al.(2017)Lu, Shanbhag, Jindal, and Madden]{Lu2017AdaptDBAP}
Y.~Lu, A.~Shanbhag, A.~Jindal, and S.~Madden, ``{{AdaptDB}}: Adaptive partitioning for distributed joins,'' \emph{Proc. of the VLDB Endowment}, vol.~10, no.~5, 2017.

\end{thebibliography}


\begin{thebibliography}{6}
\providecommand{\natexlab}[1]{#1}
\providecommand{\url}[1]{#1}
\csname url@samestyle\endcsname
\providecommand{\newblock}{\relax}
\providecommand{\bibinfo}[2]{#2}
\providecommand{\BIBentrySTDinterwordspacing}{\spaceskip=0pt\relax}
\providecommand{\BIBentryALTinterwordstretchfactor}{4}
\providecommand{\BIBentryALTinterwordspacing}{\spaceskip=\fontdimen2\font plus
\BIBentryALTinterwordstretchfactor\fontdimen3\font minus \fontdimen4\font\relax}
\providecommand{\BIBforeignlanguage}[2]{{%
\expandafter\ifx\csname l@#1\endcsname\relax
\typeout{** WARNING: IEEEtranN.bst: No hyphenation pattern has been}%
\typeout{** loaded for the language `#1'. Using the pattern for}%
\typeout{** the default language instead.}%
\else
\language=\csname l@#1\endcsname
\fi
#2}}
\providecommand{\BIBdecl}{\relax}
\BIBdecl

\bibitem[Lehmann et~al.(2023)Lehmann, Bader, Tschirpke, Thamsen, and Leser]{lehmannHowWorkflowEngines2023}
F.~Lehmann, J.~Bader, F.~Tschirpke, L.~Thamsen, and U.~Leser, ``How {{Workflow Engines Should Talk}} to {{Resource Managers}}: {{A Proposal}} for a {{Common Workflow Scheduling Interface}},'' in \emph{{Proc. CCGrid 2023}}, 2023.

\bibitem[Di~Tommaso et~al.(2017)Di~Tommaso, Chatzou, Floden, Barja, Palumbo, and Notredame]{ditommasoNextflowEnablesReproducible2017}
P.~Di~Tommaso, M.~Chatzou, E.~W. Floden, P.~P. Barja, E.~Palumbo, and C.~Notredame, ``Nextflow enables reproducible computational workflows,'' \emph{Nature Biotechnology}, vol.~35, no.~4, 2017.

\bibitem[Lehmann(2025{\natexlab{a}})]{wow_Scheduler}
\BIBentryALTinterwordspacing
F.~Lehmann, ``{Common Workflow Scheduler with WOW strategy},'' Feb. 2025. [Online]. Available: \url{https://doi.org/10.5281/zenodo.14894652}
\BIBentrySTDinterwordspacing

\bibitem[Lehmann(2025{\natexlab{b}})]{wow_Nextflow}
\BIBentryALTinterwordspacing
------, ``{Nextflow with the Common Workflow Scheduler Interface for Kubernetes extended with WOW },'' Feb. 2025. [Online]. Available: \url{https://doi.org/10.5281/zenodo.14894638}
\BIBentrySTDinterwordspacing

\bibitem[Lehmann and Tschirpke(2025)]{wow_ExperimentSetup}
\BIBentryALTinterwordspacing
F.~Lehmann and F.~Tschirpke, ``{Experiment Setup for WOW Experiments},'' Feb. 2025. [Online]. Available: \url{https://doi.org/10.5281/zenodo.14894632}
\BIBentrySTDinterwordspacing

\bibitem[Lehmann(2025{\natexlab{c}})]{wow_Results}
\BIBentryALTinterwordspacing
F.~Lehmann, ``{WOW with Nextflow and Kubernetes - Traces and Evaluation },'' Feb. 2025. [Online]. Available: \url{https://doi.org/10.5281/zenodo.14894648}
\BIBentrySTDinterwordspacing

\end{thebibliography}
\end{document}